\documentclass[a4paper,12pt]{article}
\usepackage[utf8]{inputenc}
\usepackage[UKenglish]{babel}
\usepackage{mathtools}
\usepackage{amsmath,amssymb,amsthm,amsfonts}
\usepackage[font = footnotesize]{caption}
\usepackage[ruled,vlined]{algorithm2e}
\usepackage[onehalfspacing]{setspace}
\usepackage{fancyhdr}
\usepackage{xcolor}
\usepackage{multirow}
\usepackage{float}

\usepackage[round]{natbib}
\setcitestyle{notesep={: }}
\setcitestyle{aysep={,}}

\DeclareMathOperator*{\argmin}{arg\,min}

\usepackage{titling}
\newcommand{\subtitle}[1]{%
  \posttitle{%
    \par\end{center}
    \begin{center}\large#1\end{center}
    \vskip0.5em}%
}

\usepackage[noindentafter]{titlesec}

\titlespacing\subsection{0pt}{12pt}{5pt}

\usepackage{authblk}

\renewenvironment{abstract}{%
	\vspace{6pt}%
	\begin{center}%
		\begin{minipage}{320pt}%
			\small%
			\begin{center}%
				\textbf{Abstract}%
			\end{center}%
		}{\end{minipage}\end{center}}
\newcommand{\keywords}[1]{%
	\begin{center}%
		\begin{minipage}{320pt}%
			\textit{Keywords:}~{#1}
		\end{minipage}%
	\end{center}%
}

\title{Approximate Bayesian inference for a spatial point process model exhibiting regularity and random aggregation}
\author[1]{Ninna Vihrs}
\author[1]{Jesper M\o ller}
\author[2]{Alan E. Gelfand}
\affil[1]{Department of Mathematical Sciences, Aalborg University}
\affil[2]{Department of Statistical Science, Duke University}

\begin{document}
\maketitle

\begin{abstract}
In this paper, we propose a doubly stochastic spatial point process model with both aggregation and repulsion. This model combines the ideas behind Strauss processes and log Gaussian Cox processes. The likelihood for this model is not expressible in closed form but it is easy to simulate realisations under the model. We therefore explain how to use approximate Bayesian computation (ABC) to carry out statistical inference for this model. We suggest a method for model validation based on posterior predictions and global envelopes. We illustrate the ABC procedure and model validation approach using both simulated point patterns and a real data example.
\end{abstract}

\keywords{Approximate Bayesian computation (ABC); doubly stochastic process; log Gaussian Cox process; model comparison; posterior prediction; Strauss process.}

\section{Introduction}

Spatial point patterns are usually divided into three cases: regularity/re\-pul\-sive\-ness, complete spatial randomness, and aggregation/clustering. There is a wide selection of point process models suitable for these situations, see e.g. the overview in \citet[Section 1]{clustering_regularity_thin} and the references therein. However, some point patterns show repulsiveness between the points at small scale and aggregation at a larger scale, see \cite{clustering_regularity_thin} for a detailed discussion.  In this regard, \citet{clustering_regularity_thin} suggested a model for this situation obtained by a dependent thinning of a repulsive point process. It is also possible to construct certain Gibbs point processes with this behaviour, see e.g.\ \citet{BaddeleyEtAl} and \citet{clustering_regularity_Gibbs}.

\subsection{The log Gaussian Cox Strauss process}\label{intro_model}

In this paper, we present a model for regularity at small scale and aggregation at larger scale which is a combination of a pairwise interaction point process and a log Gaussian Cox process. It is constructed by the following two steps.

First, we consider a pairwise interaction point process defined as follows. Let $\textbf{X}$ be a spatial point process viewed as a finite random subset of a given bounded region $W\subset \mathbb{R}^2$ (we think of $W$ as an observation window). Then $\textbf{X}$ is a pairwise interaction point process if $\textbf{X}$ follows a density (with respect to the unit rate Poisson process on $W$) of the form
\begin{equation}\label{eq:PIPPdensity}
f( \textbf{x} \mid\psi, \varphi) = \frac{1}{C_{\psi,\varphi}}\prod_{i=1}^n \psi(x_i)\prod_{i<j}\varphi(\|x_i-x_j\|)
\end{equation}
for all point patterns $\textbf{x}=\{x_1,\ldots,x_n\}\subset W$ with $0\le n<\infty$ (if $n=0$ then $\textbf{x}=\emptyset$ is the empty point pattern), where the notation means the following: $\psi:W\to [0, \infty)$ is a so-called first order interaction function; $\varphi:[0, \infty)\to[0, \infty)$ is a so-called second order interaction function; $\|\cdot\|$ denotes usual Euclidean distance; and $C_{\psi,\varphi}=1/f(\emptyset\mid\psi,\varphi)$ is the normalising constant which is required to be positive and finite. Usually, $\varphi(\cdot) \leq 1$, in which case the density is well defined and results in a model for repulsion between the points. The first order interaction function may be used to model systematic aggregation of points.

Second, we consider a doubly stochastic construction, by replacing $\psi$ with a random function $\Psi$ in order to introduce random aggregation to the model. This is an extension of a Cox process \citep[the case $\varphi=1$, cf.][]{cox55}, and such a model was considered in \citet{kmjm08} when $\Psi$ is the stochastic intensity function of a shot noise Cox process. Instead, we use the random intensity function of a log Gaussian Cox process \citep[\underline{LGCP}, see][]{LGCP}, which is a popular model for random aggregation. Specifically, we let
\begin{equation}\label{LGCPintensity}
\Psi(u)= \exp(Z(u)),\qquad u \in W,
\end{equation}
where $\mathbf{Z} \coloneqq \left \{Z(u)\right \}_{u\in W}$ is a Gaussian random field (\underline{GRF}) with constant mean $\mu\in\mathbb{R}$ and exponential covariance function
\begin{equation*}
c(u,v) =\sigma^2\exp \left (-\|u - v\|/s\right ),\qquad u,v\in W.
\end{equation*}
Here, $\sigma^2\ge0$ is the variance and $s>0$ is a scale parameter. For $\sigma^2>0$, the flexible stochastic process $\Psi(u)$ may account for aggregation caused by unobserved covariates. Note that  $\Psi(u)=\exp(\mu)$ if $\sigma^{2}=0$.

For the second order interaction function in \eqref{eq:PIPPdensity}, \citet{kmjm08} used a piecewise linear function, whereas we will use the much simpler second order interaction function of a Strauss process \citep{strauss1975, kelly1976}. This gives us a density for $\textbf{X}$ (with respect to the unit rate Poisson process on $W$) of the form
\begin{equation}\label{eq:density_LGCPStrauss}
f(\textbf{x} \mid\theta)= \mathrm{E}\left[\frac{1}{C_{\theta}(\textbf{Z})} \prod_{i=1}^n\exp\left (Z(x_i)\right )\prod_{i<j}\gamma^{1\left [\|x_i-x_j\|\leq R\right ]}\right],
\end{equation}
where $\theta = \left (\mu, \sigma^2, s, \gamma, R\right )$ is the parameter vector. Here, the expectation is with respect to the GRF; $C_{\theta}(\textbf{Z})$ is the normalising constant obtained by conditioning on $\textbf{Z}$; $1[\cdot]$ denotes the indicator function; and we use the convention $0^0=1$.  The parameter $R>0$ is called the interaction radius and the parameter $\gamma\in [0, 1]$ controls the repulsion between points. This model for $\textbf{X}$ will be referred to as an \underline{LGCP-Strauss}
process.

The model includes some well-known special cases:
\begin{enumerate}
\item[(a)] Conditioned on $\textbf{Z}$, $\textbf{X}$ is an inhomogeneous Strauss process.
\item[(b)] If $\sigma^2=0$, $\textbf{X}$ is a usual Strauss process. If in addition $\gamma = 0$, $\textbf{X}$ is a hard core Gibbs process with hard core parameter $R$; or if in addition $\gamma=1$, $\textbf{X}$ is a homogeneous Poisson process on $W$ with intensity $\exp(\mu)$.
\item[(c)] If $\gamma = 1$, $\textbf{X}$ is an LGCP.
\end{enumerate}

The following coupling result becomes useful when interpreting the meaning of $\gamma$ and when we later discuss simulation of the LGCP-Strauss process. To stress the dependence on $\gamma$, we write $\textbf{X}=\textbf{X}_\gamma$. Then, using a dependent thinning technique   \citep{KendallMoeller} it follows that there exists a coupling of the LGCP-Strauss processes $\textbf{X}_\gamma$ for all $\gamma\in[0,1]$ such that $\textbf{X}_\gamma\subseteq \textbf{X}_{\gamma'}$ whenever $0\le\gamma<\gamma'\le1$. In particular, the special case of the LGCP $\textbf{X}_1$ (item (c) above) dominates any of the LGCP-Strauss processes $\textbf{X}_\gamma$. The intensity of $\textbf{X}_1$ is $\exp(\mu+\sigma^2/2)$ \citep{LGCP}, so $\exp(\mu+\sigma^2/2)|W|$ provides an upper bound on the expected number of points in $\textbf{X}_{\gamma}$. Here, $|W|$ denotes the area of $W$.

Note that if we are not in any of the above special cases (a)--(c), both the intensity and other moment characteristics of $\textbf{X}$, the density \eqref{eq:density_LGCPStrauss}, and the Papangelou conditional intensity \citep[see e.g.][]{textbook} are not expressible in closed form. Therefore, in general, usual approaches for estimation based on likelihood, pseudo-likelihood, composite likelihood, and minimum contrasts \citep[see the review in][]{JMRW2017} are not feasible for the LGCP-Strauss process. This makes statistical inference challenging. Finally, note that for a Poisson process `everything is known', whilst for a Strauss process the Papangelou conditional intensity but not the moment characteristics are expressible in closed form, and for an LGCP the moment characteristics but not the Papangelou conditional intensity are expressible in closed form, cf.\ the above-mentioned references.

\subsection{Objective and outline}

In this paper, we show how to use approximate Bayesian computation (ABC) to make statistical inference for spatial point process models such as the LGCP-Strauss process model. In brief, ABC is a flexible method for approximate inference in a Bayesian framework, which does not require the likelihood to be expressible in closed form. Instead, it is based on the ability to make simulations under the assumed model, which are then compared to the observed data by using summary statistics.

In previous work on ABC in the setting of spatial point process models, \citet{ABCppAlan} explained how ABC can be used for Strauss process models and determinantal point process models. For the Strauss process model they estimated the interaction radius using maximum profile pseudo likelihood and then kept the interaction radius fixed at this estimate during the ABC procedure. Further, \citet{ABCppfunctional} presented an ABC method using functional summary statistics such as the pair correlation function, which they exemplified for a Thomas process model and a marked point process model. Finally, \citet{ABCshadow} presented an ABC method for spatial point process models dealing with an intractable normalising constant in the likelihood. This method will not help for the LGCP-Strauss process since it is not only a normalising constant but also the expectation in \eqref{eq:density_LGCPStrauss} which makes the density intractable.

In contrast to \citet{ABCppAlan}, the method we use for statistical inference is based entirely on ABC, and unlike \citet{ABCppAlan} and \citet{ABCppfunctional} we do not fix any of the unknown parameters during the ABC procedure. Furthermore, we provide a discussion of the choice of summary statistics for ABC when making statistical inference for the LGCP-Strauss process. We also suggest a method for model validation and comparison based on posterior predictions and global envelopes. We use this in a simulation study to assess the quality of ABC results for LGCP-Strauss processes and to investigate whether realisations of the LGCP-Strauss process can be distinguished from LGCPs and Strauss processes.

The remainder of this paper is organized as follows. Section~\ref{sec:simex} presents simulated examples of LGCP-Strauss processes. In Section~\ref{sec:ABCpp}, our chosen method for ABC model fitting is specified.
Section~\ref{sec:ABCsim} contains ABC analyses for simulated data. Section \ref{sec:ABCoak} contains a real data example using a point pattern of oak trees which suffer from frost shake.  Section~\ref{s:final} concludes with a brief summary and paths for future work.

The open source software R \citep{R} is used for all statistical computations. Most plots are created with the R-package \texttt{ggplot2} \citep{ggplot} and some of the functionalities of the R-package \texttt{spatstat} \citep{spatstat} are used to handle spatial point patterns.

\section{Simulation study of the LGCP-Strauss process}
\label{sec:simex}
Consider an LGCP-Strauss process $\textbf{X}$ on the observation window $W$ with density \eqref{eq:density_LGCPStrauss}, which depends on the parameter vector $\theta =(\mu,\sigma^{2},s,\gamma, R)$. We simulate data under this model in two steps: First, a realisation $\textbf{z}$ of $\textbf{Z}$ is simulated (see e.g.\ \citet{schlather99}). In R, this can be done with the function \texttt{RFsimulate} from the R-package \texttt{RandomFields} \citep{RF2,RF1}. Second, a realisation of $\textbf{X}$ given $\textbf{Z}=\textbf{z}$ is simulated using an MCMC algorithm, namely a birth-death Me\-tro\-po\-lis-Ha\-stings algorithm \citep[][specifically, a birth is proposed with probability $1/2$ and otherwise a death is proposed; for a birth proposal, the new point is generated from a density proportional to $\exp\left (\textbf{z} \right )$; and for a death proposal, the point to die is selected uniformly from the current point pattern]{bdMetropolis}.

\begin{figure}[ht!]
\centering
\includegraphics[scale = 0.3]{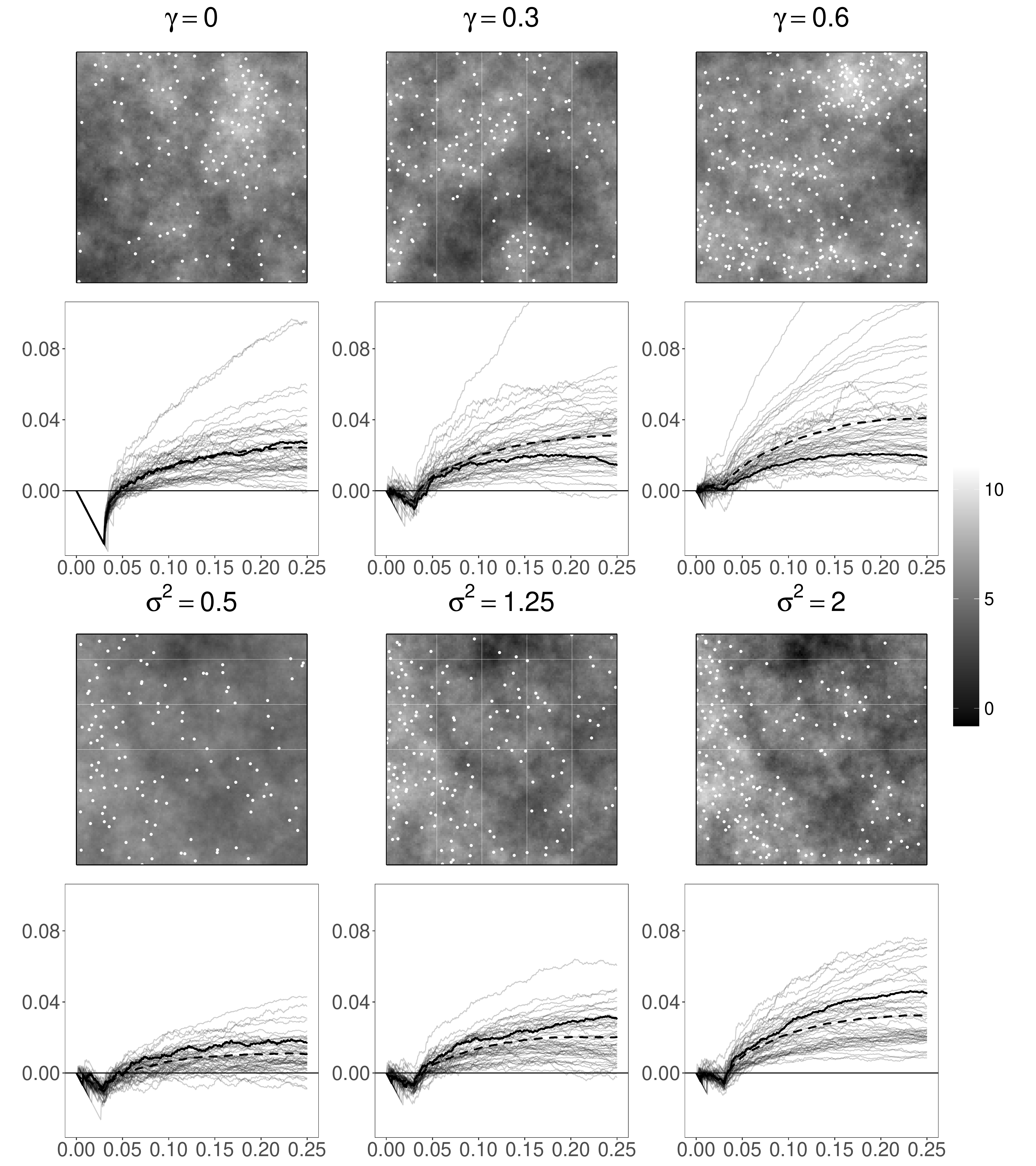}
\caption{First and third row: simulated LGCP-Strauss processes on the unit square (white points) and the corresponding realisation of $\textbf{Z}$ (grey scale image). In the first row, the parameters are $\mu = 5,~\sigma^2=2,~R=0.03$, $s = 0.3$, and $\gamma$ is as specified at the top of each plot. In the third row, the parameters are $\mu = 5,R=0.03$, $s = 0.2, \gamma = 0.3$, and $\sigma^2$ is as specified at the top of each plot. Second and fourth row: empirical $L$-function minus the identity for the point pattern directly above the plot (solid curve) and for $49$ different simulations of the same process (grey curves) plus their mean (dashed curve).}
\label{fig:sim_ex}
\end{figure}

Figure \ref{fig:sim_ex} shows six examples of simulated realisations of the LGCP-Strauss process on the unit square (using a burnin of $20\,000$ in the MCMC algorithm) plotted on top of the corresponding realisation of $\textbf{Z}$. The processes generating the first three point patterns only differ by the value of $\gamma$ and the ones generating the last three only differ by the value of $\sigma^2$; the remaining parameters are specified in the caption.

To asses the degree of clustering and regularity we consider the $L$-function $L(r)=\sqrt{K(r)/\pi}$, where $r>0$ denotes inter-point distance and $K$ is Ripley's $K$-function \citep{ripley76, ripley77}. The $L$-function is commonly used to summarise important aspects of the second order moment properties of spatial point processes. Since $L(r)=r$ for a Poisson process, one usually considers $T(r)\coloneqq L(r)-r$. The $L$-function is often used to make statements about clustering/regularity as follows: If $L(r)<r$ ($L(r)>r$), this indicates that $\textbf{X}$ is regular/repulsive (aggregated/clustered) at inter-point distances $r$ (for more detailed explanations, see e.g.\ \citet{spatstat}).

Figure~\ref{fig:sim_ex} also shows plots of empirical estimates of the $L$-functions of the point patterns using Ripley's isotropic edge correction \citep{ripley77}(an alternative to displaying the empirical $L$-functions would show the empirical pair correlation functions though they are sensitive to choice of bandwidth). Each of these plots also includes the empirical $L$-functions of 49 different realisations of the process in question plus the mean of these in order to assess the general behaviour of the estimator (note that the mean does not necessarily represent the behaviour of the theoretical $L$-function because the estimator is biased). As expected, the point patterns exhibit both regularity and aggregation. The first three point patterns show a decreasing degree of regularity at small to moderate distances as $\gamma$ increases, but a similar degree of aggregation at large distances. However, the general behaviour of the empirical $L$-function suggests a tendency to a higher degree of clustering at large distances as $\gamma$ increases. The last three point patterns show a similar degree of regularity at small to moderate distances and an increasing degree of clustering at large distances as $\sigma^2$ increases. We also see that the variance of the estimator of the $L$-function at large distances apparently increases as $\sigma^2$ increases. Note that the empirical $L$-function of the first point pattern, which is generated from the LGCP-Strauss process where $\sigma^2=0.5$, would not be uncommon for realisations of the LGCP-Strauss processes where $\sigma^2 = 1.25$ or $\sigma^2=2$ either. This suggest that it may be difficult to see the effect of $\sigma^2$ on the clustering when looking at a given realisation. Notice that in general the repulsive behaviour of the point patterns to some extent obscures the finer variations in $Z$ (especially for strong repulsion), so overall we may expect that it will be difficult to make inference for the parameters of the GRF.

\section{ABC for spatial point process models}\label{sec:ABCpp}
ABC is a method used to make approximate Bayesian inference for a parametric model with an intractable likelihood by developing an approximate posterior sample of the parameters. Instead of having to evaluate the likelihood, it is only necessary to be able to simulate from the model in order to do ABC. The most basic ABC technique is ABC rejection sampling which goes as follows: for a parametric model with parameter $\theta$ and observed data $x_{\mathrm{obs}}$, specify a summary statistic $S$ and a distance function $\chi$; repeat
sampling $\theta'$ from its prior, and $x'$ given $\theta'$ from the likelihood, until $\chi(S(x'),S(x_{\mathrm{obs}}))<\varepsilon$ where $\varepsilon$ is a chosen tolerance; then return $\theta'$. If the inequality $\chi(S(x'),S(x_{\mathrm{obs}}))<\varepsilon$ is replaced by $S(x')=S(x_{\mathrm{obs}})$ and $S$ is either the identity function or a sufficient statistic, $\theta'$ will be an exact sample from the posterior distribution. When this ideal situation is not achievable, the sample will instead be from an approximation to the posterior distribution, referred to as the ABC posterior. The quality of this approximation will depend on the choice of $\varepsilon$ and $S$. There exist many different ABC techniques and some even circumvent the need for summary statistics, see e.g. \citet{ABCKullbackLeibler}, \citet{ABCMMD}, and \citet{ABCWasserstein}. However, it does not seem to be straightforward to apply these methods to the setting of spatial point process models.

In this paper, we illustrate how ABC can be used to make inference for the LGCP-Strauss process.  A relatively simple ABC method is successful for this illustration; it is specified in Section~\ref{sec:ABCspecification}.  So, we need not discuss or compare different ABC techniques.  For a more detailed overview of ABC and some of the different techniques, see e.g. \citet{ABCoverview}.

\subsection{Specification of the ABC procedure}
\label{sec:ABCspecification}

Consider a spatial point process $\textbf{X}$ defined on a bounded region $W\subset \mathbb{R}^{2}$ and which follows a parametric model with parameter vector $\theta$. Assume a realisation $\textbf{x}_{\mathrm{obs}}$ of $\textbf{X}$ is observed. Our chosen procedure for ABC is specified in Algorithm \ref{algo:ABC_procedure} below. It is inspired by \citet{ABCppAlan} and the semi-automatic approach by \citet{ABCsemi}. \citet{ABCppAlan} used a Markov chain Monte Carlo method for the ABC sampling whereas we choose ABC rejection sampling, because of its simplicity and ability to be run in parallel. The semi-automatic part refers to the fact that the user specified summary statistics are only used in a pilot run instead of directly in the ABC rejection step.

{\small
\begin{algorithm}[htp!]
\SetKwInOut{Input}{Input}\SetKwInOut{Output}{Output}
\SetKwFor{For}{For}{}{endfor}
\DontPrintSemicolon
\Input{Data $\textbf{x}_{\mathrm{obs}}$, a prior distribution $\pi(\theta)$ for $\theta=(\theta_{1}, \ldots, \theta_{p})$, a procedure for simulating from the likelihood $\pi(\textbf{x}\mid\theta)$, a summary statistic $T(\textbf{x})=(T_1(\textbf{x}),\ldots, T_d(\textbf{x}))$, positive integers $k_{\mathrm{pilot}}$ and $k_{\mathrm{ABC}}$, and a non-negative integer $m$.}
\Output{A sample $\theta^{\mathrm{ABC},1},\ldots,\theta^{\mathrm{ABC},k_{\mathrm{ABC}}}$ from the ABC approximate posterior distribution.}
Calculate $T_{\mathrm{obs}}=T(\textbf{x}_{\mathrm{obs}})$.

\textbf{Pilot run}:

\For{$i=1,\ldots,k_{\mathrm{pilot}}$}{
\Repeat{$n(\mathbf{x}^{\mathrm{pilot},i})>m$.}{
sample $\theta^{\mathrm{pilot}, i}\sim\pi(\theta)$ and $\textbf{x}^{\mathrm{pilot},i}\sim\pi(\textbf{x}\mid\theta^{\mathrm{pilot}, i})$
}
}
\For{$j=1,\ldots,p$}{
based on the sample $\left \{\left (\theta^{\mathrm{pilot},i}, \textbf{x}^{\mathrm{pilot},i}\right )\right \}_{i=1}^{k_{\mathrm{pilot}}}$, fit a linear model for the posterior mean $$\mathrm{E}[\theta_j\mid \textbf{x}]\approx\theta_j(\textbf{x})\coloneqq\alpha^j+\beta^{j^T} (T(\textbf{x}) - T_{\mathrm{obs}})$$ where $\textbf{x}$ is a realisation of $\textbf{X}$, $\alpha^j\in\mathbb{R}$, and $\beta^j=(\beta_{1}^{j},\ldots,\beta_{d}^j)\in\mathbb{R}^{d}$. Let $\hat{\theta}_j(\textbf{x})$ be the estimate of $\theta_{j}(\textbf{x})$ when $\alpha^j$ and $\beta^j$ are replaced by the estimates $\hat{\alpha}^j$ and $\hat{\beta}^j$.

}
Define the distance measure
\begin{equation*}
\chi\left (\hat\theta(\textbf{x}),\hat\theta(\textbf{x}_{\mathrm{obs}})\right )=\sum_{j=1}^{p}\frac{\left (\hat{\theta}_j\left (\textbf{x}\right )-\hat{\theta}_j\left (\textbf{x}_{\mathrm{obs}}\right )\right )^2}{\hat{\text{var}}\left (\hat{\theta}_j\right )} = \sum_{j=1}^{p}\frac{\left (\hat{\theta}_j\left (\textbf{x}\right )-\hat{\alpha}^j\right )^2}{\hat{\text{var}}\left (\hat{\theta}_j\right )}
\end{equation*}
where $\hat\theta(\textbf{x}) = (\hat{\theta}_1\left (\textbf{x}\right ),\ldots,\hat{\theta}_p\left (\textbf{x}\right ))$ and $\hat{\text{var}}\left (\hat{\theta}_j\right )$ is the empirical variance of $\left \{\hat{\theta}_{j}(\textbf{x}^{\mathrm{pilot},i})\right \}_{i=1}^{k_{\mathrm{pilot}}}$.\;
Choose $\varepsilon$ as the empirical $1\%$ percentile of $\left \{\chi(\textbf{x}^{\mathrm{pilot},i},\textbf{x}_{\mathrm{obs}})\right \}_{i=1}^{k_{\mathrm{pilot}}}$.

\textbf{ABC rejection sampling}:

\For{$i=1,\ldots,k_{\mathrm{ABC}}$}{
\Repeat{$\chi(\hat\theta(\mathbf{x}^{\mathrm{ABC},i}),\hat\theta(\mathbf{x}_{\mathrm{obs}}))<\varepsilon$}{
\Repeat{$n(\mathbf{x}^{\mathrm{ABC},i})>m$.}{
sample $\theta^{\mathrm{ABC}, i}\sim\pi(\theta)$ and $\textbf{x}^{\mathrm{ABC},i}\sim\pi(\textbf{x}\mid\theta^{\mathrm{ABC}, i})$
}
}
}

\caption{Procedure for ABC}
\label{algo:ABC_procedure}
\end{algorithm}}

In Algorithm \ref{algo:ABC_procedure}, $n(\textbf{x})$ is the number of points in a point pattern $\textbf{x}$, and in the first and last for loop we demand that $n(\textbf{x})>m$ for each simulated $\textbf{x}$. This is not strictly necessary for ABC, but it is a way to insure that summary statistics are not calculated for point patterns with very few points. Most summary statistics for spatial point patterns can only be calculated or considered reliable if there is a reasonable number of points in the point pattern. In the examples of Sections \ref{sec:ABCsim} and \ref{sec:ABCoak}, $m = 10$ was found to be sufficient.

In the second for loop of Algorithm \ref{algo:ABC_procedure}, we choose to fit the linear models approximating the posterior means $\mathrm{E}[\theta_i\mid \textbf{x}]$, $i=1,\ldots,p$, with a special case of a relaxed Lasso \citep{relaxedlasso}: First, a model is fitted with Lasso regression, where the penalty term is chosen based on a cross-validation argument using the `one-standard-error rule' \citep[see e.g.][Chapter 2]{lasso}. Let $\hat{\beta}_j^{i,\mathrm{Lasso}}$, $j = 1,\ldots, d$, be the resulting estimate of $\beta_{j}^{i}$ and set $T^{i,\mathrm{Lasso}}(\textbf{x})=\left \{T_j(\textbf{x})\mid\hat{\beta}_j^{i,\mathrm{Lasso}}\neq 0, j=1,\ldots, d\right \}$. Second, the summary statistics in $T^{i,\mathrm{Lasso}}$ are used as predictors in a linear model fitted with ordinary least squares, which results in the final model for approximating $\mathrm{E}[\theta_i\mid\textbf{x}]$. We employ a Lasso regression approach because we want to use a relatively large number of summary statistics (see Section~\ref{sec:ABC_LGCPStrauss}).

\subsection{Choice of summary statistics in the case of LGCP-Strauss process models}\label{sec:ABC_LGCPStrauss}

An important part of ABC is the selection of appropriate summary statistics. It is not possible to find sufficient statistics for the LGCP-Strauss process, since it does not have a closed form density and, therefore, it is not obvious which summary statistics to use. We emphasize that there are limitless possibilities for choosing such summary statistics.  The following describes one choice which, based on some theoretical knowledge and numerical experiments, we believe is an appropriate set of summary statistics when implementing ABC for LGCP-Strauss processes.

Recall from Section~\ref{intro_model} that $\exp(\mu + \sigma^2/2)|W|$ provides an upper bound on the expected number of points for an LGCP-Strauss process on $|W|$. We may therefore expect that especially the parameters $\mu$ and $\sigma^2$ are related to the number of points in a point pattern generated by an LGCP-Strauss process. We therefore include the number of observed points as a summary statistic.

Recall also the $L$-function from Section~\ref{sec:simex} which is a theoretical tool commonly used to asses the degree of clustering and regularity. Since these properties are related to many of the parameters of the LGCP-Strauss process, we consider an empirical estimate of the $L$-function among the summary statistics for ABC (see (b)-(c) below). A simulation study  suggested that for realisations of an LGCP-Strauss process, the empirical  estimate of $L(r)-r$ often has a global minimum when $r$ is close to the interaction radius $R$, at least when there is strong to moderate repulsion in the model. In this regard, see Figure~\ref{fig:sim_ex} for some examples of empirical $L$-functions associated with realisations of LGCP-Strauss processes. We take this into consideration when choosing the summary statistics (see (b) below).

Furthermore, numerical experiments suggested that it may be particularly difficult to learn much about the GRF based on a realisation (see also the discussion in Section~\ref{sec:simex}). The GRF mainly affects the clustering, so we would like to include some further summary statistics which can capture this. For this purpose, assume for ease of exposition that $W$ is a square with side length $h$. Then we split $W$ into $q^2$ squares $W_{i,j}$ of side length $h/q$, $i,j=1,\ldots,q$, and let $n(\textbf{x}\cap W_{i,j})$ be the number of points in $\textbf{x}$ falling in $W_{i,j}$. We choose summary statistics which describe how $n(\textbf{x}\cap W_{i,j})$ varies (see (d) below) and which are calculated for a user-specified finite range of $q$-values.

Specifically, for a point pattern $\textbf{x}$ (either $\textbf{x}_{\mathrm{obs}}$ or one of the simulated point patterns in Algorithm \ref{algo:ABC_procedure}), we chose the following summary statistics.
\begin{enumerate}
\item[(a)] $n_{\mathrm{log}}\coloneqq\log(n(\textbf{x}))$.
\item[(b)] $L_{\mathrm{max}}\coloneqq\max(\hat{L}(r)-r)$,\\ $L_{\mathrm{min}}\coloneqq\min(\hat{L}(r)-r)$, and\\ $L_{\mathrm{arg\,min}}\coloneqq\argmin(\hat{L}(r)-r)$,\\ where $\hat{L}$ is a non-parametric estimate of the $L$-function evaluated over a user-specified finite range of $r$-values.
\item[(c)] $\hat{L}(r)-r$ evaluated  at $m$ equally spaced values of $r$ between 0 and $0.2h$ referred to as $L_1,\ldots, L_m$.
\item[(d)] $C_{\mathrm{max}, q}\coloneqq\underset{i,j=1,\ldots,q}{\max}\left (\left \{ n(\textbf{x}\cap W_{i,j})/n(\textbf{x})\right\}\right )$,\\$C_{\mathrm{min},q}\coloneqq\underset{i,j=1,\ldots,q}{\min}\left (\left \{ n(\textbf{x}\cap W_{i,j})/n(\textbf{x})\right\}\right )$, and \\$C_{\mathrm{log\,var},q}\coloneqq\log\left (\hat{\text{var}}\left (\left \{ n(\textbf{x}\cap W_{i,j})/n(\textbf{x})\right\}_{i,j=1}^{q}\right )\right )$, \\where again $\hat{\text{var}}$ means empirical variance.
\end{enumerate}
We have chosen these specific forms of the summary statistics based on some numerical experiments. In the examples of Sections \ref{sec:ABCsim} and \ref{sec:ABCoak}, $m = 40$ and $q=2,\ldots, 5$. This means that the vector of summary statistics $T$ in Algorithm~\ref{algo:ABC_procedure} has dimension equal to $1 + 3 + 40 +3\cdot 4= 56$.

\section{ABC for simulated realisations of LGCP-Strauss processes}\label{sec:ABCsim}

\subsection{Prior specification and numerical considerations}\label{sec:sim_data_priors}
We will now illustrate how the procedure in Algorithm~\ref{algo:ABC_procedure} can be used to make ABC for the simulated realisations of LGCP-Strauss processes in Figure~\ref{algo:ABC_procedure}. In order to do this, it is required to specify a (proper) prior distribution for the parameter vector $\theta=(\mu,\sigma^2,s,\gamma,R)$ of the LGCP-Strauss process. For the examples in this section, we considered three different prior distributions for $\theta$ which we refer to as  $P_1$, $P_2$, and $P_3$. In each case, a priori we assume the five parameters $\mu$, $\sigma^2$, $s$, $\gamma$, and $R$ are independent.
\begin{enumerate}
\item $P_1$: $\mu\sim\mathrm{Unif}(3, 6)$, $\sigma^2\sim\mathrm{Unif}(0, 4)$, $s\sim\mathrm{Unif}(0.01, 0.5)$, $\gamma\sim\mathrm{Unif}(0,1)$, and $R\sim\mathrm{Unif}(0, 0.05)$;
\item $P_2$: $\mu\sim\mathrm{Norm}(3.5, 1)$, $\sigma^2\sim\mathrm{Gamma}(1,1)$, $s\sim\mathrm{Gamma}(1, 6)$,\\ $\gamma\sim\mathrm{Beta}(1,2)$, and $R\sim\mathrm{Gamma}(1, 50)$;
\item $P_3$: $\mu\sim\mathrm{Norm}(5, 1)$, $\sigma^2\sim\mathrm{Gamma}(10,4)$, $s\sim\mathrm{Gamma}(7, 20)$,\\ $\gamma\sim\mathrm{Beta}(2,1)$, and $R\sim\mathrm{Gamma}(10, 250)$.
\end{enumerate}
Here, $\mathrm{Unif}(a, b)$ is the uniform distribution on the interval $(a, b)$, $\mathrm{Norm}(a,b)$ is the normal distribution with mean $a$ and variance $b$, $\mathrm{Gamma}(a, b)$ is the gamma distribution with shape $a$ and rate $b$, and $\mathrm{Beta}(a, b)$ is the beta distribution with the shape parameters $a$ and $b$. For computational reasons, we have chosen to truncate all the prior distributions for the parameters to the intervals of the uniform distributions of $P_1$. For example, the computational reasons include a consideration of the number of points in simulations. The more points a simulated point pattern has, the more computationally expensive the simulation procedure will be (see below). Recall that $\exp(\mu+\sigma^2/2)$ is an upper bound on the expected number of points in the unit square. By limiting the range of $\mu$ and $\sigma^2$ in their prior distributions, we ensure that point patterns simulated during the ABC procedure will not yield unreasonably many points compared to the number of points in our observed point patterns.  Further, this ensures that the simulated point patterns can achieve regularity similar to that in the observed point patterns.

In order to use the MCMC algorithm when a realisation $\textbf{z}=\{z(u)\}_{u\in W}$ of the GRF is given (see Section \ref{sec:simex}), it is necessary to choose a burn-in which can be used for all simulations in the ABC procedure. In order to choose this burn-in, we considered 30 samples of the parameters drawn from the prior distribution $P_1$; used the MCMC algorithm for all these samples; and considered trace plots of the number of points and $R$-close pairs. Figure~\ref{fig:trace_plots} shows these plots for three different prior samples for illustration. We choose to initiate the MCMC algorithm at the empty point pattern and at a realisation of an inhomogeneous Poisson process on $W$ with intensity function $\exp(z(u))$ (these initial states are extreme because of the coupling result mentioned in Section~\ref{intro_model}). It seems that the higher the number of points, the slower the convergence. The burn-in should be high enough for the MCMC algorithm to have converged given any prior sample, but increasing the burn-in will also increase the computation time. Considering all 30 examples, $20\,000$ appears to be an appropriate overall burn-in. All following simulations are iteration $20\,001$ of the MCMC algorithm initiated at the empty point pattern.

\subsection{Posterior results}\label{sec:ABCposteriors}

\begin{figure}[htp!]
\centering
\includegraphics[scale=0.33]{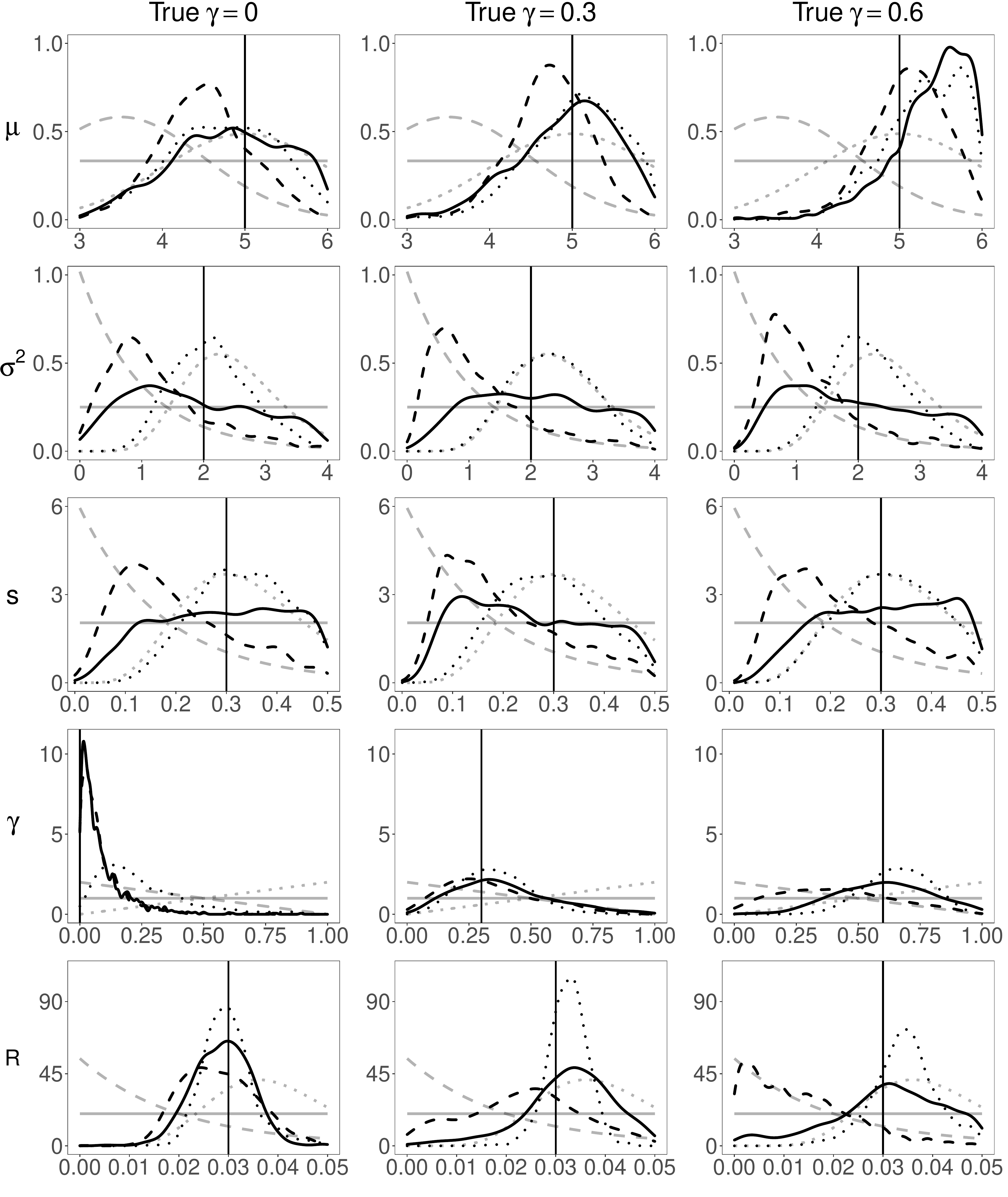}
\caption{Estimated marginal ABC posterior densities (black curves) when using different prior distributions (grey curves) for the parameters of the LGCP-Strauss model used for the first three point patterns in Figure \ref{fig:sim_ex}. The vertical lines indicate the true parameter values. For each marginal ABC posterior distribution, the corresponding prior distribution is plotted using the same linetype (solid for $P_1$, dashed for $P_2$, and dotted for $P_3$). Each row represents a parameter (stated to the left of the row), and each column represents one of the three point patterns, as indicated by the true value of $\gamma$.}
\label{fig:ABC_marginals_sim}
\end{figure}

\begin{figure}[htp!]
\centering
\includegraphics[scale=0.33]{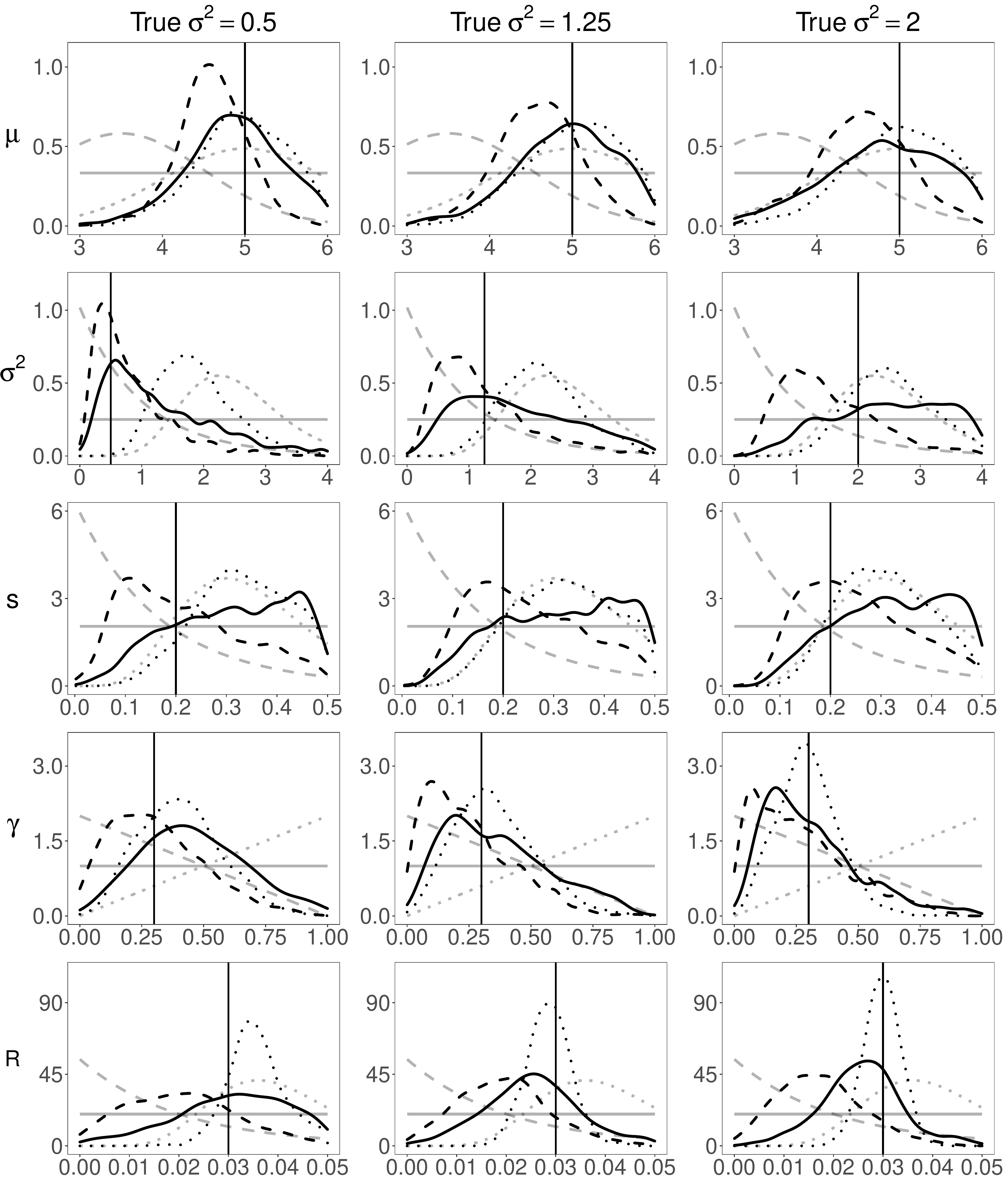}
\caption{Estimated marginal ABC posterior densities (black curves) when using different prior distributions (grey curves) for the parameters of the LGCP-Strauss model used for the last three point patterns in Figure \ref{fig:sim_ex}. The vertical line indicates the true parameter value. For each posterior distribution, the corresponding prior distribution is plotted using the same linetype (solid for $P_1$, dashed for $P_2$, and dotted for $P_3$). Each row represents a parameter (stated to the left of the row), and each column represents one of the three point patterns, as indicated by the true value of $\sigma^2$.}
\label{fig:ABC_marginals_sim2}
\end{figure}

We used Algorithm \ref{algo:ABC_procedure} on the six point patterns in Figure \ref{fig:sim_ex} with $k_{\mathrm{pilot}}=10\,000$  and $k_{\mathrm{ABC}}=1\,000$ \citep[the same choice as in][]{ABCppAlan}. In some cases, one or two pilot samples had to be excluded afterwards because some summary statistics yielded infinite values. For a single point pattern, in our situation, it usually took about 10 hours to run the ABC procedure in parallel on 45 cores (evidently, run time will depend heavily on the given situation and software). Figures~\ref{fig:ABC_marginals_sim} and \ref{fig:ABC_marginals_sim2} show kernel density estimates of the resulting (approximate) marginal posterior distributions of the parameters, using a Gaussian kernel and a bandwidth chosen with the method by \citet{bwselection}.

From Figure~\ref{fig:ABC_marginals_sim}, we see the following.
\begin{itemize}
\item As the true value of $\gamma$ increases, the ABC posterior distributions of $\mu$ become more and more left skewed. The choice of prior seems to have small influence on the general behaviour of these. The ABC posteriors corresponding to the prior $P_2$ are very different from the prior in all three cases, whereas the ABC posteriors corresponding to the priors $P_1$ and $P_3$ seem to become increasingly different from their corresponding priors as $\gamma$ increases.
\item The ABC posterior distributions for $\sigma ^2$ and $s$ look rather similar to their prior distributions, except near zero in the situations of the priors $P_1$ and $P_2$ where the ABC posteriors are considerably smaller than their corresponding priors. This suggests that even though it may be difficult to infer with precision about the values of $\sigma^2$ and $s$, we are able to learn that they have a small probability of being near zero, which means that we can detect a clustering effect in the point patterns.
\item When the true value of $\gamma$ is 0, the ABC posteriors of $\gamma$ are very concentrated near 0. We see a tendency for the spread of the ABC posteriors to increase as $\gamma$ increases. The choice of prior seems to have small influence on the overall behaviour of the ABC posteriors in the first two cases. Especially for the prior $P_1$, the maxima of the ABC posteriors seem to be in good agreement with the true value in all three cases.
\item The posterior distributions for $R$ seem to approach the corresponding priors as $\gamma$ increases. In all three cases, the maxima of the posterior distributions corresponding to the prior $P_1$ are in good agreement with the true value.
\end{itemize}

From Figure~\ref{fig:ABC_marginals_sim2}, we see the following.
\begin{itemize}
\item
The overall behaviour of the ABC posteriors for $\mu$ seems to be rather unaffected by the choice of prior, and the ABC posteriors seem to be in good agreement with the true value. The spread of these ABC posteriors seems to increase slightly as $\sigma^2$ increases.
\item For $\sigma^2$, the ABC posteriors corresponding to the prior $P_3$ are quite similar to the prior. The posteriors corresponding to the prior $P_1$ are getting closer to the prior as the true value of $\sigma^2$ increases. The spread of the posteriors corresponding to the priors $P_1$ and $P_2$ seems to be increasing as $\sigma^2$ increases. Again, we see that the posteriors corresponding to the priors $P_1$ and $P_2$ place less mass near zero than the corresponding priors.
\item For $s$, the results are very similar to those in Figure~\ref{fig:ABC_marginals_sim}.
\item For $\gamma$, the choice of prior seems to have little influence on the ABC posteriors, and overall the spread of the ABC posteriors seems to decrease slightly as $\sigma^2$ increases.
\item For $R$, the prior seems to have some influence on the ABC posteriors and the spread of the ABC posteriors seems to be decreasing when $\sigma^2$ is increasing.
\end{itemize}

\begin{figure}[ht!]
\centering
\includegraphics[scale=0.33]{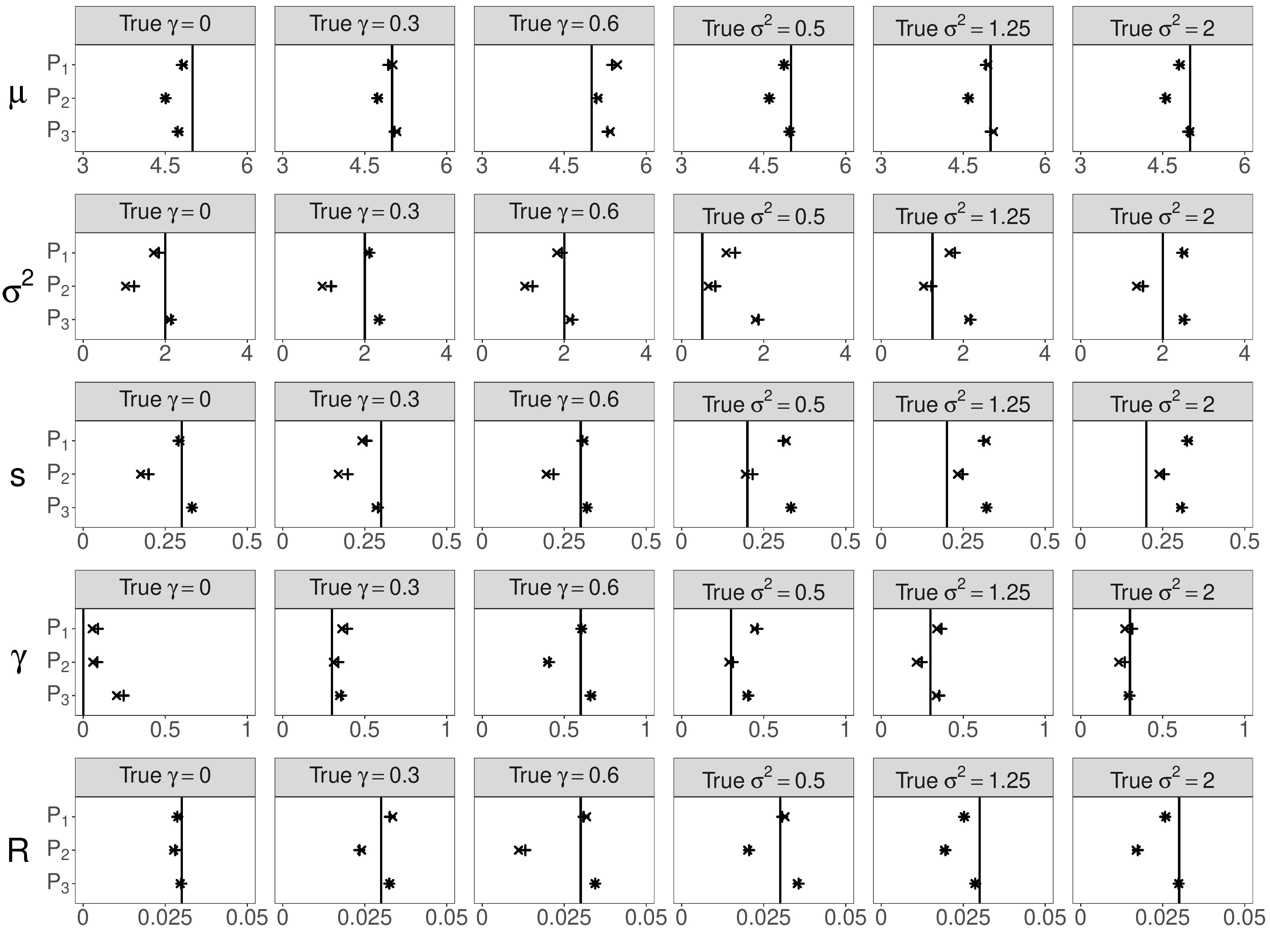}
\caption{Posterior means (indicated by +) and medians (indicated by x) of the ABC samples for the point patterns in Figure \ref{fig:sim_ex} (which are referred to by the true value of $\gamma$ for the first three and the true value of $\sigma^2$ for the last three). Each row represent a parameter, which is stated to the left of the row. The prior distribution is stated on the $y$-axis and the $x$-axis correspond to the full range of each parameter.}
\label{fig:post_means_medians}
\end{figure}

Figure~\ref{fig:post_means_medians} shows the means and medians of the ABC samples, where in all cases the posterior mean and median are close. Furthermore, we see the following:
\begin{itemize}
\item
 For $\mu$, all the posterior means and medians for the different priors and point patterns look similar except in the case $\gamma = 0.6$ where they are somewhat higher than in the other cases. Overall, they are in fairly good agreement with the true values.
 \item For $\sigma^2$ and $s$, we again see that it is quite difficult to obtain much precision about these parameters from data.
 \item For $\gamma$, the prior has the smallest influence when the true value of $\gamma$ is relatively low and the true value of $\sigma^2$ is relatively high, in which case the ABC posterior means and medians are also very close to the true value.
 \item For $R$, the prior has the smallest influence when the true value of $\gamma$ is small. Considering the priors $P_1$ and $P_3$, the means and medians are generally close to the true value. For the prior $P_2$, the means and medians seem to become increasingly smaller than the true value of $R$ as the true value of $\gamma$ increases.
 \end{itemize}

Overall, the ABC procedure seems to be most successful for estimating $\mu$, $\gamma$ and $R$, especially when the true value of $\gamma$ is relatively small and the value of $\sigma^2$ is not too small. However, the success of the procedure will depend on the specific combination of the true parameters. Note that when fitting a Strauss process to a point pattern, \citet{ABCppAlan} first estimated $R$ by maximum pseudo-likelihood and then used this value of $R$ in their ABC procedure; in contrast, we found no need to fix $R$ when fitting an LGCP-Strauss process with our ABC procedure.

\citet{covar_unidentifiable} demonstrated that some of the parameters
in the Mat{\'e}rn model (which includes the exponential covariance function) may not be consistently identified in an increasing density
infill asymptotics framework, but that the parameter $\sigma^2/s$ may
be consistently identified.
This might explain why the ABC procedure is not so successful when it comes to identifying the scale and variance parameters of an exponential covariance function. Therefore, we made the same analysis as in Figure~\ref{fig:ABC_marginals_sim} using the prior $P_1$ when $s=0.3$ is given. However, the posterior marginal distributions of the remaining parameters (not shown) looked very similar to those in Figure \ref{fig:ABC_marginals_sim}.

Table~\ref{tab:coef_linmod} summarises the estimated coefficients of the linear models fitted in the ABC procedure, cf. Algorithm~\ref{algo:ABC_procedure}. For each parameter, there are a total of 18 fitted linear models (one for each time the ABC procedure was run, that is, one for each combination of point pattern and prior). The table shows for each parameter the mean (over all 18 linear models) of the estimated coefficient for each summary statistic. In order to make the coefficients for different summary statistics comparable in this table, the linear models were fitted to a scaled version of the training data (the scaling was done by subtracting the mean and dividing by the standard deviation). As expected, $n_{\mathrm{log}}$ is most influential in the models for $\mu$ and $\sigma^2$. Of the summary statistics of the type $C_{\mathrm{log\,var}, q}$, $C_{\mathrm{min}, q}$, and $C_{\mathrm{max}, q}$, it appears that $C_{\mathrm{log\,var}, 5}$ is the most important one. For all parameters, some values of the empirical $L$-function seem to have some influence in the linear models. The summary statistic $L_\mathrm{min}$ is most influential in the models for $\gamma$, which is in agreement with the fact that it describes the degree of regularity. Interestingly, $L_{\mathrm{arg\,min}}$ does not appear to be very influential in the models related to $R$, in contrast to what might be expected.

\begin{table}[htp!]
\scriptsize
\centering
\begin{tabular}{l|ccccc}
  \hline
Summary statistic & $\gamma$ & $\mu$ & $R$ & $s$ & $\sigma^2$ \\
  \hline
$n_{\mathrm{log}}$ & -0.001 & 0.662 & -0.168 & 0.05 & 0.337 \\
$C_{\mathrm{log\,var},2}$ & -0.043 & 0.005 & - & 0.046 & - \\
$C_{\mathrm{log\,var},3}$& - & - & - & 0.179 & - \\
$C_{\mathrm{log\,var},4}$& 0.035 & -0.002 & -0.046 & 0.046 & 0.088 \\
$C_{\mathrm{log\,var},5}$& 0.225 & -0.133 & -0.179 & -0.017 & 0.171 \\
$C_{\mathrm{min},2}$& - & -0.008 & - & -0.08 & -0.009 \\
$C_{\mathrm{min},3}$& -0.004 & 0.001 & -0.009 & -0.036 & -0.061 \\
$C_{\mathrm{min},4}$& -0.015 & 0.058 & 0.004 & - & -0.029 \\
$C_{\mathrm{min},5}$& 0.029 & 0.076 & -0.033 & 0.038 & -0.027 \\
$C_{\mathrm{max},2}$& 0.008 & - & - & 0.006 & -0.057 \\
$C_{\mathrm{max},3}$& 0.023 & 0.025 & - & - & -0.058 \\
$C_{\mathrm{max},4}$& 0.031 & - & -0.02 & - & -0.077 \\
$C_{\mathrm{max},5}$& - & 0.036 & - & - & -0.146 \\
$L_2$ & 0.093 & -0.032 & 0.052 & -0.067 & 0.054 \\
$L_3$ & 0.064 & -0.018 & 0.021 & -0.039 & 0.053 \\
$L_4$ & 0.055 & -0.005 & 0.035 & -0.044 & 0.014 \\
$L_5$ & 0.059 & -0.026 & 0.039 & -0.024 & 0.023 \\
$L_6$ & 0.091 & -0.011 & 0.058 & -0.024 & 0.007 \\
$L_7$ & 0.117 & -0.007 & -0.035 & -0.004 & -0.013 \\
$L_8$ & 0.072 & -0.074 & -0.145 & -0.031 & -0.075 \\
$L_9$ & 0.041 & -0.007 & -0.201 & - & -0.032 \\
$L_{10}$ & 0.044 & -0.058 & -0.388 & -0.044 & -0.064 \\
$L_{11}$ & - & - & -0.008 & -0.014 & - \\
$L_{12}$ & -0.039 & - & - & -0.022 & 0.022 \\
$L_{13}$ & -0.036 & - & - & -0.064 & 0.045 \\
$L_{14}$ & -0.045 & -0.001 & - & -0.055 & 0.06 \\
$L_{15}$ & -0.011 & - & 0.015 & -0.14 & 0.078 \\
$L_{16}$ & - & - & - & -0.006 & 0.052 \\
$L_{17}$ & -0.027 & - & 0.103 & - & 0.023 \\
$L_{18}$ & - & - & 0.009 & -0.008 & 0.089 \\
$L_{19}$ & -0.01 & - & - & - & 0.062 \\
$L_{20}$ & -0.049 & - & 0.023 & - & 0.044 \\
$L_{21}$ & -0.023 & - & - & - & 0.008 \\
$L_{23}$ & -0.089 & - & 0.003 & - & 0.026 \\
$L_{24}$ & - & - & 0.022 & - & 0.026 \\
$L_{26}$ & -0.011 & - & 0.056 & - & - \\
$L_{27}$ & -0.009 & - & 0.008 & - & - \\
$L_{28}$ & - & 0.01 & 0.074 & - & 0.039 \\
$L_{29}$ & -0.01 & 0.027 & -0.024 & - & - \\
$L_{30}$ & - & 0.002 & 0.039 & - & - \\
$L_{31}$ & -0.036 & - & - & - & - \\
$L_{32}$ & - & -0.002 & - & - & - \\
$L_{33}$ & 0.023 & - & - & - & - \\
$L_{34}$ & -0.043 & -0.002 & -0.001 & - & - \\
$L_{35}$ & -0.104 & - & 0.02 & - & - \\
$L_{37}$ & - & - & 0.034 & - & - \\
$L_{38}$ & 0.019 & - & - & - & - \\
$L_{39}$ & - & 0.005 & 0.016 & 0.01 & - \\
$L_{40}$ & -0.021 & -0.018 & 0.04 & 0.148 & - \\
$L_{\mathrm{arg\,min}}$ & -0.087 & -0.01 & 0.009 & 0.001 & 0.041 \\
$L_{\mathrm{min}}$ & 0.12 & 0.027 & 0.017 & 0.085 & -0.085 \\
$L_{\mathrm{max}}$ & - & - & 0.035 & 0.161 & -0.005 \\
   \hline
\end{tabular}
\caption{Table of mean of estimated coefficients in the linear models fitted in the ABC procedure for each parameter. The data was scaled before the models were fitted in order to make the coefficients comparable. Summary statistics whose coefficients were zero in all models are not included.}
\label{tab:coef_linmod}
\end{table}

We also investigated the estimated intercept (on the original scale) of the linear models for each parameter, point pattern, and prior. According to the model specification in Algorithm~\ref{algo:ABC_procedure}, the intercept represents a linear approximation to the expected value of the parameter given the observed point pattern. Overall, these estimated intercepts were similar to the ABC posterior means in Figure~\ref{fig:post_means_medians} and are therefore not shown.

\begin{figure}[H]
\centering
\includegraphics[scale=0.3]{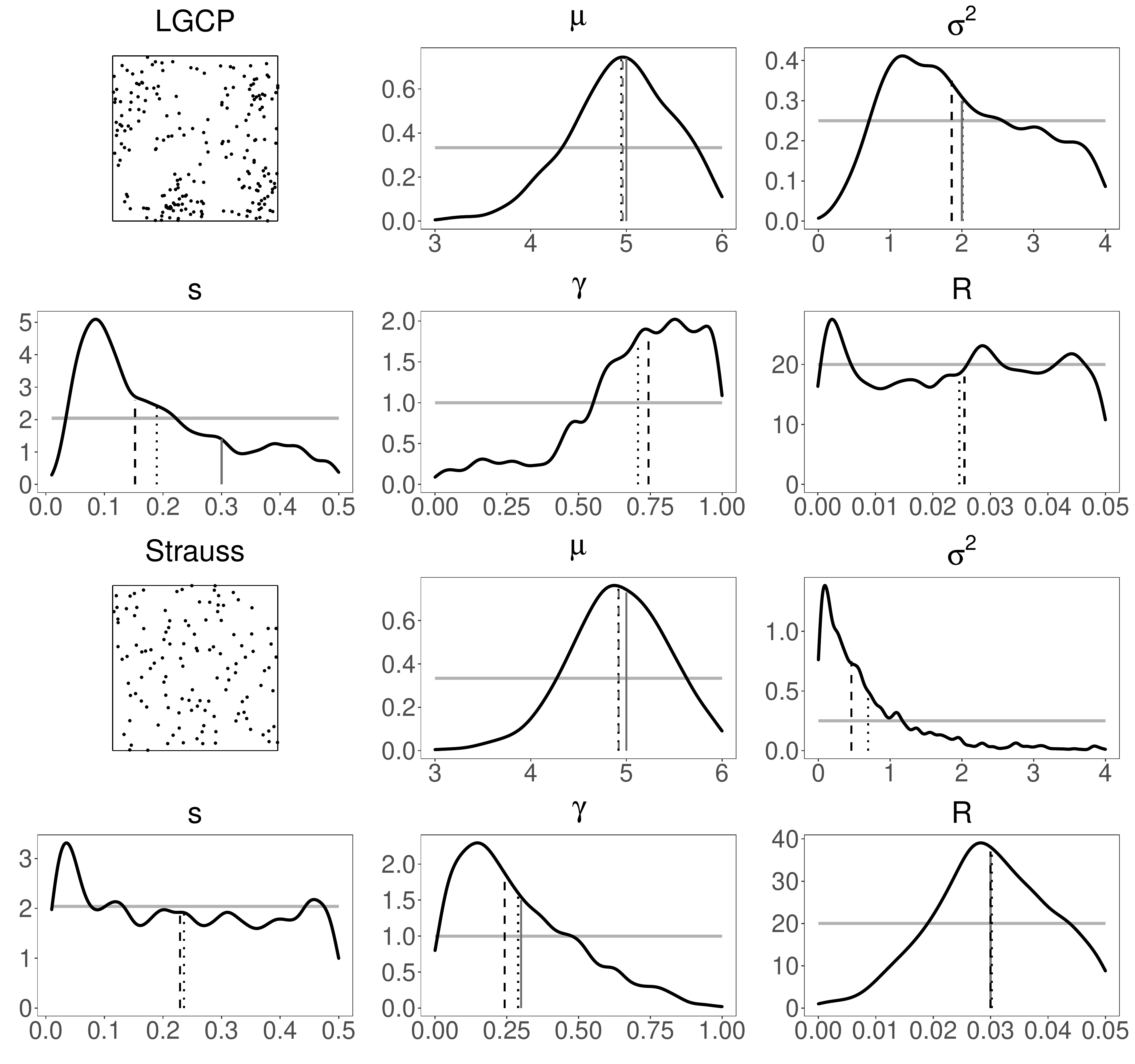}
\caption{Panels 2-6 and 8-12 show estimated marginal posterior distributions (black curves) and corresponding prior distributions (grey curves) for the parameters (as stated on the top) when fitting an LGCP-Strauss process to the realization of an LGCP in panel 1 (corresponding to panels 2-6) and the realization of a Strauss process in panel 7 (corresponding to panels 8-12). The point processes are defined on the unit square. The dashed and dotted vertical lines indicate the medians and means, respectively; the solid vertical lines indicate the true values, when relevant.}
\label{fig:ABCspecial}
\end{figure}

We now investigate how the ABC procedure for fitting an LGCP-Strauss process works when the data is generated from some of the special cases of this process. For this purpose, we simulated a realisation of an LGCP with parameters $\mu = 5,~\sigma^{2}=2$, and $s=0.3$, and a realisation of a Strauss process with parameters $\mu = 5,~\gamma = 0.3$, and $R=0.03$. Notice that when simulating under an LGCP, there is no need to employ the MCMC algorithm described at the beginning of Section~\ref{sec:simex}. We used the faster method implemented in the function \texttt{rLGCP} from the package \texttt{spatstat} \citep{spatstat},  which meant that we were able to run this ABC procedure for a single point pattern in about 40 minutes (again using 45 cores). We used the same ABC procedure as above with the specified uniform priors for fitting an LGCP-Strauss process to these point patterns and the posterior results can be seen in Figure~\ref{fig:ABCspecial}. For the point pattern generated from an LGCP, the true value of $\mu$ seems to be identified well when fitting the LGCP-Strauss process. The posterior marginal distribution of $\gamma$ is rather concentrated near $1$, and a plot of the posterior samples of $\gamma$ and $R$ (not shown) shows that very small values of $\gamma$ appear together with very small values of $R$. This indicates that the fitted LGCP-Strauss process is close to the special case of an LGCP, which is the true model. Again, it seems to be difficult to identify $\sigma^{2}$ and $s$.

For the point pattern generated from a Strauss process, the marginal ABC posterior distribution for $\sigma^{2}$ is very concentrated near zero, which is the true value. The true values of $\mu$ and $R$ seems to be well identified, and the median and mean of the marginal posterior distribution of $\gamma$ are quite close to the true value, but the maximum value of this ABC posterior distribution is somewhat smaller than the true value. For the Strauss process, $s$ should be irrelevant, which is in agreement with the nearly uniform ABC posterior distribution for $s$.

\subsection{Model checking and comparison}
\label{sec:model_checking}
We are interested in whether the point patterns in Figure~\ref{fig:sim_ex} can be distinguished from realisations of an LGCP and a Strauss process, so for comparison we also fitted an LGCP and a Strauss process to each point pattern, using the ABC procedure in Algorithm~\ref{algo:ABC_procedure}. We used the same summary statistics as for the LGCP-Strauss process and the prior $P_1$ specified in \ref{sec:sim_data_priors} on the relevant parameters (that is, the parameters $\mu,~\sigma^2$, and $s$ when fitting the LGCP, and the parameters $\mu,~\gamma$, and $R$ when fitting the Strauss process). Again, when simulating under an LGCP, we used the faster method implemented in \texttt{spatstat}.

For model checking and comparison we first suggest to make global envelope tests based on posterior predictions as follows. For each ABC realisation of $\theta$, a realisation $\textbf{x}$ of the process in question given $\theta$ is simulated. For each $\textbf{x}$, a functional summary statistic is estimated. These empirical curves are then used to construct global envelopes and corresponding tests based on extreme rank lengths \citep[note that we only used 1\,000 simulations instead of the recommended 2499, because the ABC procedure is rather time consuming]{GET2017, GET2018}. The R-package \texttt{GET} \citep{GET2017} was used for this purpose.

In order to compare the fitted LGCP-Strauss, LGCP, and Strauss process models, we used $95\%$ global envelopes based on posterior predictions and the empirical $L$- and $J$-function, with $J(r)=(1-G(r))/(1-F(r))$ where $F$ is the empty space function and $G$ is the nearest-neighbour distribution function \citep[see][]{Jfunction}. We also tried to use the $F$- and $G$-functions for model validation but these functional summary statistics were unable to distinguish between the models (just reflecting the well-known fact that the $J$-function contains other information than each of the $F$- and $G$-functions).

\begin{figure}[ht!]
\centering
\includegraphics[scale=0.33]{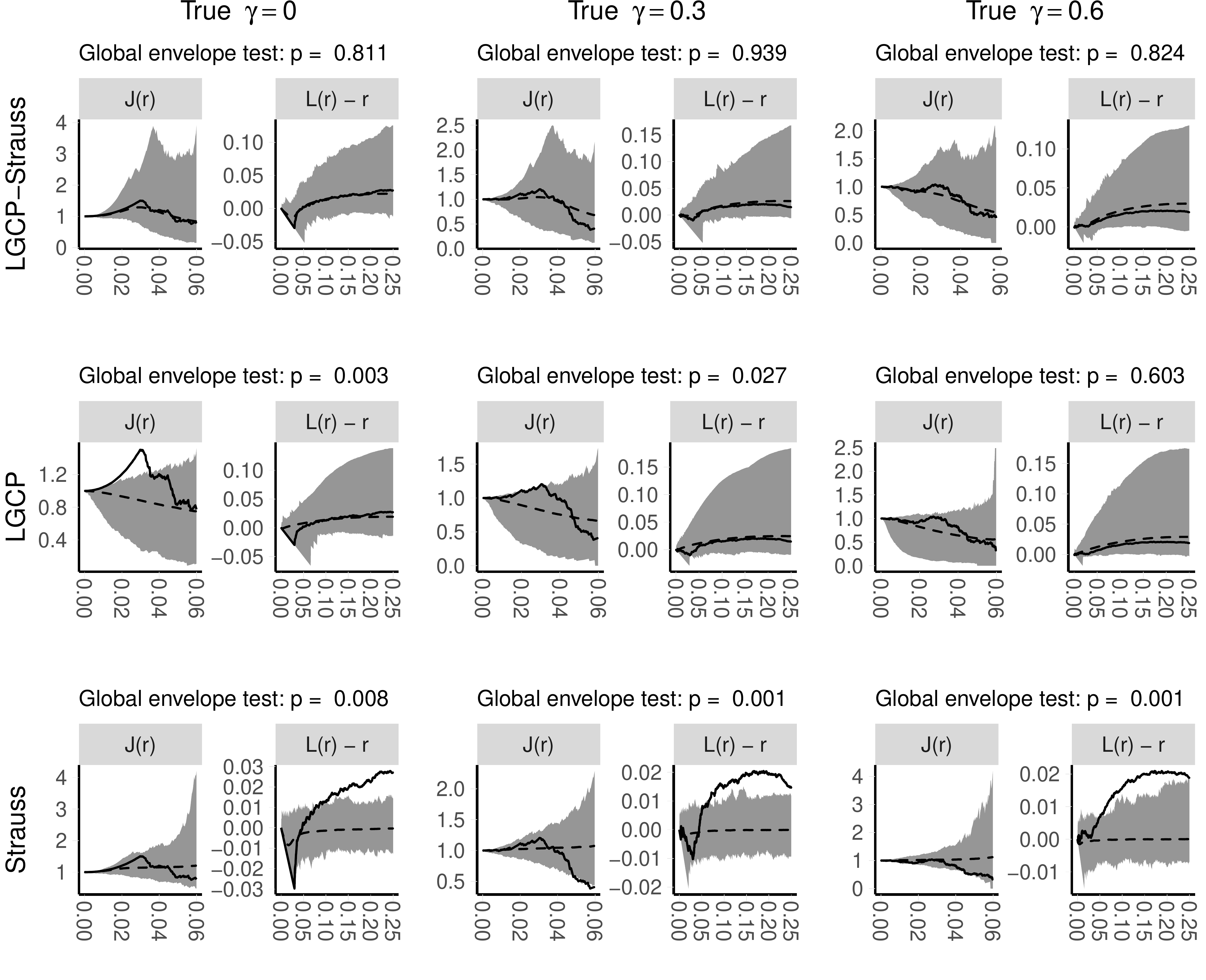}
\caption{Combined global envelopes based on the empirical $J$- and $L$-function for LGCP-Strauss, LGCP, and Strauss processes fitted with ABC to the three point patterns in Figure \ref{fig:sim_ex}. The choice of the fitted model is stated to the left of each row and each column represents a different point pattern, as indicated by the true value of $\gamma$. The solid curves are the empirical functional summary statistics for the observed point patterns and the dashed curves are the means obtained from 1\,000 posterior predictions. Each shaded area indicates a $95\%$ global envelope based on the extreme rank length. At the top of each plot, the $p$-value of the corresponding global envelope test is stated. }
\label{fig:env_comb_sims}
\end{figure}

\begin{figure}[ht!]
\centering
\includegraphics[scale=0.33]{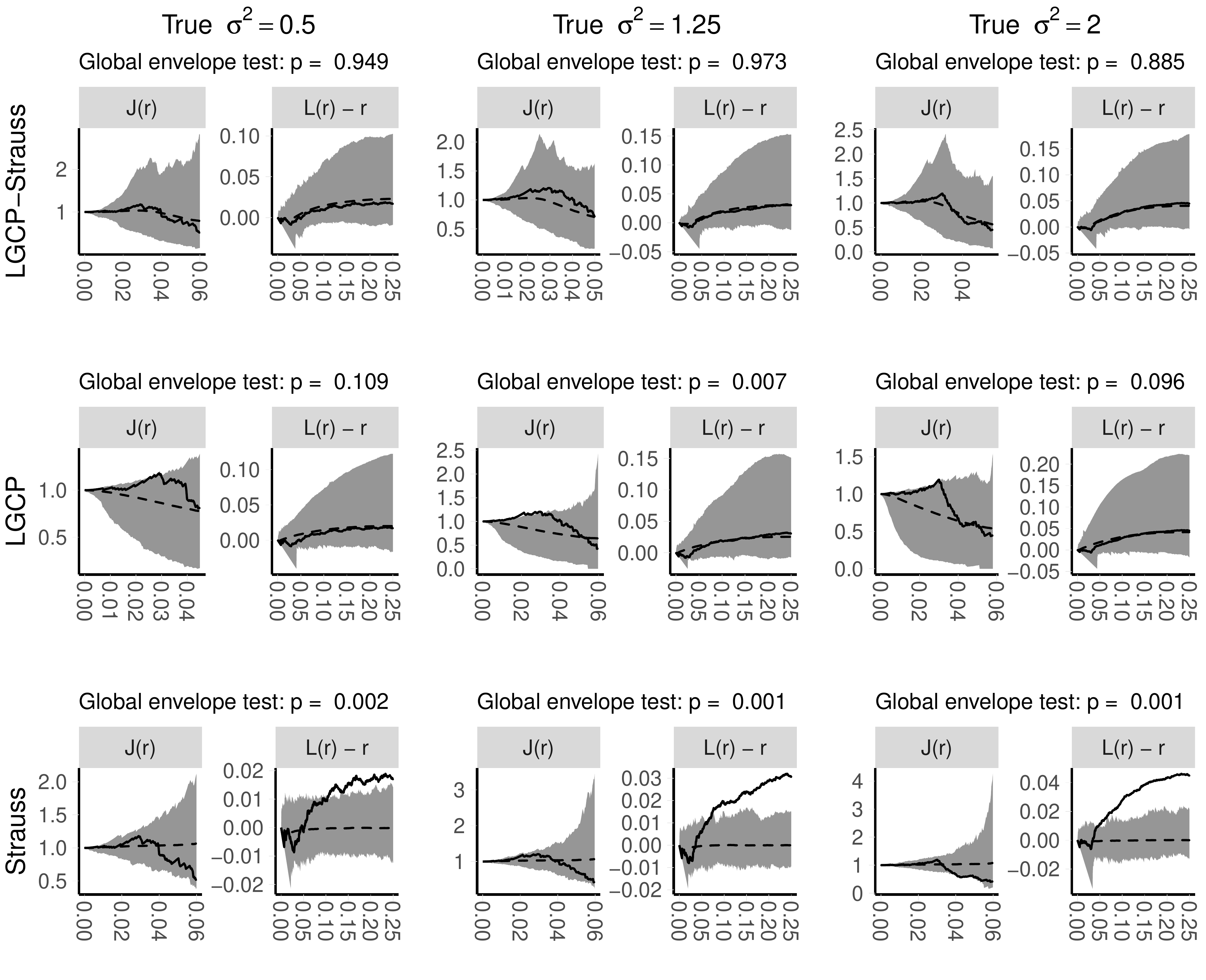}
\caption{Combined global envelopes based on the empirical $J$- and $L$-function for LGCP-Strauss, LGCP, and Strauss processes fitted with ABC to the last three point patterns in Figure \ref{fig:sim_ex}. The choice of the fitted model is stated to the left of each row and each column represents a different point pattern, as indicated by the true value of $\sigma^2$. The solid curves are the empirical functional summary statistics for the observed point patterns and the dashed curves are the means obtained from 1\,000 posterior predictions. Each shaded area indicates a $95\%$ global envelope based on the extreme rank length. At the top of each plot, the $p$-value of the corresponding global envelope test is stated. }
\label{fig:env_comb_sims2}
\end{figure}

Figures~\ref{fig:env_comb_sims} and \ref{fig:env_comb_sims2} show $95\%$ combined global envelopes for the $L$- and $J$-function, meaning that, under the LGCP-Strauss process, the probability that both empirical curves are within their respective envelopes is approximately $95\%$. To combine the envelopes we have used the two-step combining procedure described in \citet{GETinR}. The posterior predictions of the LGCP-Strauss processes are for the ABC samples from Section~\ref{sec:ABCposteriors} obtained with the prior $P_1$. Note that the $J$-function can only be estimated reliably for all simulations for $r$-values in a relatively small interval, whereas the $L$-function can be estimated reliably on a larger interval.

In all cases, the $p$-values of the global envelope tests are highest in the situation of the LGCP-Strauss process, which may indicate that they provide the best fit to data. Considering Figure \ref{fig:env_comb_sims}, the LGCP is rejected in the cases where $\gamma = 0$ and $\gamma = 0.3$ because the empirical $J$-functions in these cases are above the $95\%$ global envelopes at small inter-point distances.  This indicates that the point patterns are more regular at small inter-point distances than what would be expected under the fitted LGCPs. For the case $\gamma = 0.6$ (the case with weakest inhibition), the LGCP cannot be rejected. Notice that the $p$-values of these tests are increasing as $\gamma$ increases which is in agreement with the fact that the LGCP-Strauss process approaches the special case of an LGCP. Considering Figure \ref{fig:env_comb_sims2}, the LGCP is only rejected in the case where $\sigma ^ 2 = 1.25$, but the $p$-values are also rather small in the other two situations. In all three situations, the behaviour of the empirical $J$-function indicates that the point patterns are somewhat more regular than what is typical under the fitted LGCP.

The Strauss process model is rejected in all six cases because the empirical $L$-function clearly shows that the point patterns are more clustered at moderate to large inter-point distances than what can be modelled with a Strauss process. In  the cases $\gamma = 0.3$ in Figure \ref{fig:env_comb_sims} and $\sigma^2 = 2$ in Figure \ref{fig:env_comb_sims2}, the empirical $J$-functions also show this, but for the remaining cases, the $J$-function is contained completely within the envelopes.

Overall, it appears that the $J$-function is best at criticizing the LGCP and the $L$-function is best at criticizing the Strauss process. The later may have something to do with the fact that the $J$-function can only be estimated on a relatively small interval.  So, it is less likely to capture the aggregation, which happens on a larger scale, than the $L$-function which can be estimated on a bigger interval. When we use the $L$-function for model validation we keep in mind that it was also used in the ABC procedure which might lead us to conclude that the model fits better to data than it actually does.

The global envelope tests are mainly a method for model validation, but they may be used for model comparison by comparing $p$-values and concluding that the model with the highest $p$-value provides the best fit. However, it should be kept in mind that a higher $p$-value may be a result of overfitting.

In order to investigate this, we also fitted an LGCP to the first point pattern in Figure~\ref{fig:ABCspecial} and a Strauss process model to the second point pattern in Figure~\ref{fig:ABCspecial} and compare them to the fitted LGCP-Strauss process models (the global envelopes are not shown). For the realisation of an LGCP, the $p$-values of the $95\%$ combined global envelope test for the fitted LGCP-Strauss and LGCP were $0.933$ and $0.766$, respectively. Since the data is generated from an LGCP, both models should fit the data equally well, so the higher $p$-value for the LGCP-Strauss process is probably a result of the fact that it is overfitting. For the realisation of a Strauss process model, the $p$-values of the $95\%$ combined global envelope test for the fitted LGCP-Strauss and Strauss process were $0.46$ and $0.766$, respectively. In this example, the $p$-values do not reveal the fact that the LGCP-Strauss process is overfitting.

We also consider an ABC method for model comparison, using the method of ABC model choice via random forests (ABC-RF) and the corresponding R-package \texttt{abcrf} from \citet{ABCRF}. In short, the idea is to make a number of prior predictions (including a model index); calculate summary statistics for these; create a reference table with model indices and calculated summary statistics; and finally use this table to train a random forest classifier for predicting the model from the summary statistics. This classifier is then used on the summary statistics of the observed data to choose a model. \citet{ABCRF} also described how to approximate the posterior probability of the chosen model. According to \citet{ABCRF}, the method is robust to the number and choice of summary statistics.

We used this method for the point patterns in Figures~\ref{fig:sim_ex} and \ref{fig:ABCspecial}, using a uniform prior on the three models in consideration (LGCP-Strauss, LGCP, and Strauss), the uniform priors of $P_1$ for the relevant parameters, the summary statistics from Section~\ref{sec:ABC_LGCPStrauss}, and $30\,000$ prior predictions (whereof eight were afterwards excluded because some summary statistics yielded infinite values). The number of prior predictions are in agreement with the recommendations in \citet{ABCRF}. We used the default settings from the \texttt{abcrf} package for the remaining choices concerning the ABC-RF method.

The results are in Table~\ref{tab:ABCRF_sims}. The true model is chosen in all cases, and the approximate posterior probabilities are very high in the cases where the true model is LGCP-Strauss or Strauss. When the true model is an LGCP, the approximate posterior probability is somewhat smaller.

\begin{table}[ht!]
\begin{tabular}{llc}
Point pattern & Selected model & Approximate posterior probability\\ \hline
True $\gamma = 0$ & LGCP-Strauss & 0.97\\
True $\gamma = 0.3$ & LGCP-Strauss & 0.99\\
True $\gamma = 0.6$ & LGCP-Strauss & 0.88\\
True $\sigma^2 = 0.5$ & LGCP-Strauss & 0.96\\
True $\sigma^2 = 1.25$ & LGCP-Strauss & 0.98\\
True $\sigma^2 = 2$ & LGCP-Strauss & 0.99\\
True LGCP & LGCP & 0.65\\
True Strauss & Strauss & 0.94
\end{tabular}
\caption{Selected model and its approximate posterior probability when using ABC-RF for the point patterns in Figures~\ref{fig:sim_ex} and \ref{fig:ABCspecial}.}
\label{tab:ABCRF_sims}

\end{table}

Note that this method is intended for choosing between different types of models.  Whether a model of the chosen type actually fits the data is assessed by the global envelope test. So, a model choice method such as ABC-RF may then be particularly useful for choosing between different types of models which according to the global envelopes all fit the data.

\section{Data example}\label{sec:ABCoak}

\begin{figure}[ht!]
\centering
\includegraphics[scale=0.3]{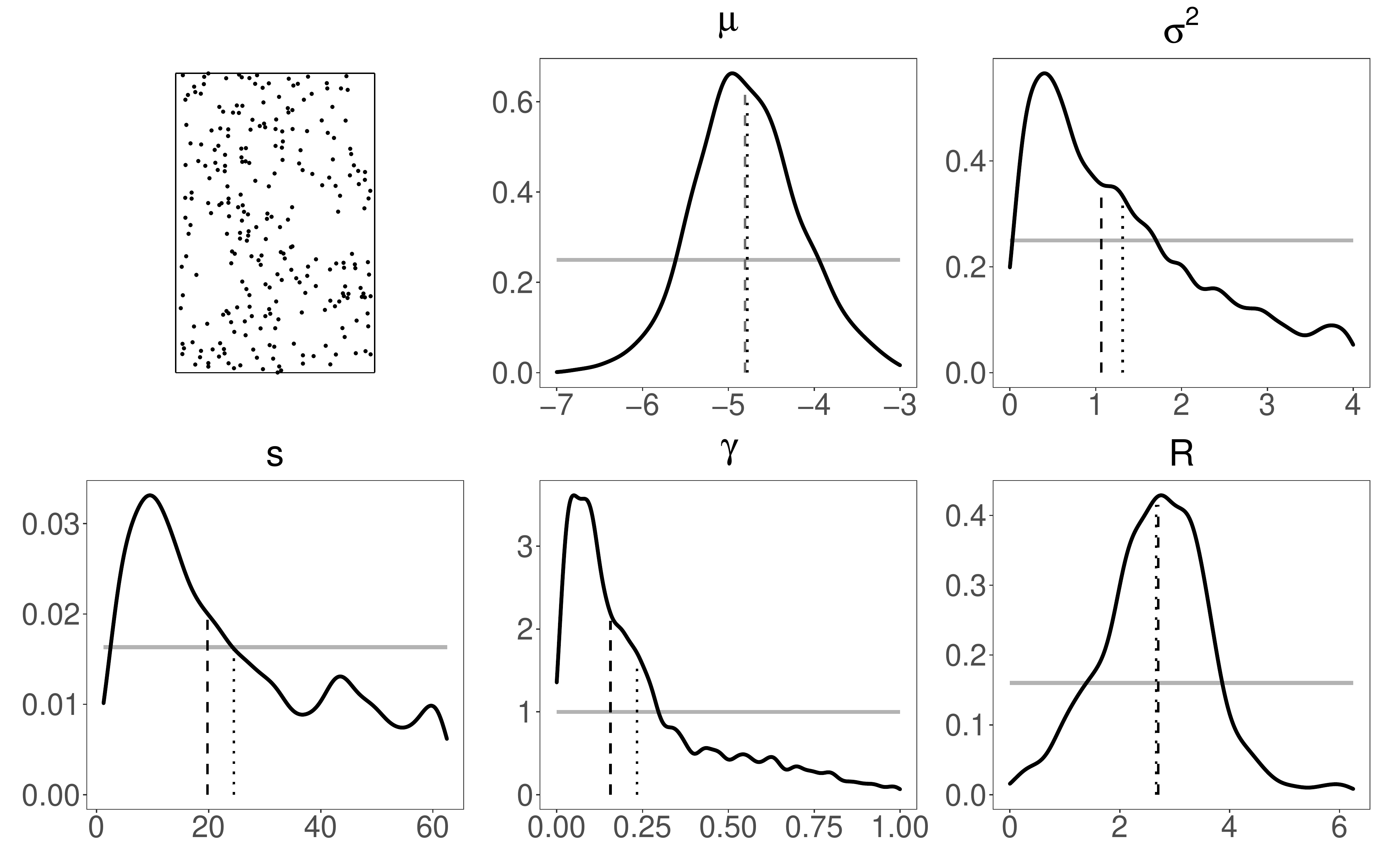}
\caption{The first panel shows the frost shake oak point pattern dataset where the observation window is a $125\times 188$ m rectangle. The other panels show the estimated marginal ABC posterior distributions (black curves) and the prior distributions (grey curves) for the five parameters, with each parameter stated at the top of each plot. The dashed and dotted lines indicate the medians and means, respectively.}
\label{fig:ABCoak}
\end{figure}

The first panel in Figure \ref{fig:ABCoak} shows the locations of 256 oak trees which suffer from frost shake (frost shake refers to cracks in the trunk of the tree) in a $125\times 188$ m rectangular region of Allogny in France. This data set is part of the Allogony data set from the R-package \texttt{ads} \citep{ads}.

We used Algorithm \ref{algo:ABC_procedure} on this oak data set.
Here, independent uniform prior distributions are chosen for $\mu$ on the interval $(-7, -3)$, $\sigma^{2}$ on $(0, 4)$, $s$ on $(1.25, 62.5)$, $\gamma$ on $(0, 1)$, and $R$ on $(0, 6.25)$. Notice that the observation window for the oak data is much larger than the ones in Section~\ref{sec:sim_data_priors}, and the prior distributions are chosen to take this into account. Furthermore, when calculating the summary statistics for the ABC procedure, $W_{i,j}$, $i,j =1,\ldots,q$, are now rectangular sets of the same size (see Section~\ref{sec:ABC_LGCPStrauss}). Trace plots as those in Figure \ref{fig:trace_plots} (supplied in an appendix) suggested that $20\,000$ iterations of the MCMC algorithm is a sufficient burn-in for this example. Again, a pilot sample of $10\,000$ simulations was used and the resulting ABC posterior sample consists of $1\,000$ draws from the approximate posterior distribution.

The marginal posterior distributions, which are estimated from the ABC sample, can be seen in Figure \ref{fig:ABCoak}. They are all clearly different from their uniform priors.
The posterior distributions of $\mu$ and $R$ look approximately normal, whilst the posterior distributions of $\sigma^2,~s$, and $\gamma$ are right skew. Note that the posterior distribution of $\gamma$ indicates strong repulsion between the points. The posterior distribution of $\sigma^{2}$, particularly its heavy tail, suggests some aggregation among the splited oaks.

\begin{figure}[ht!]
\centering
\includegraphics[scale=0.33]{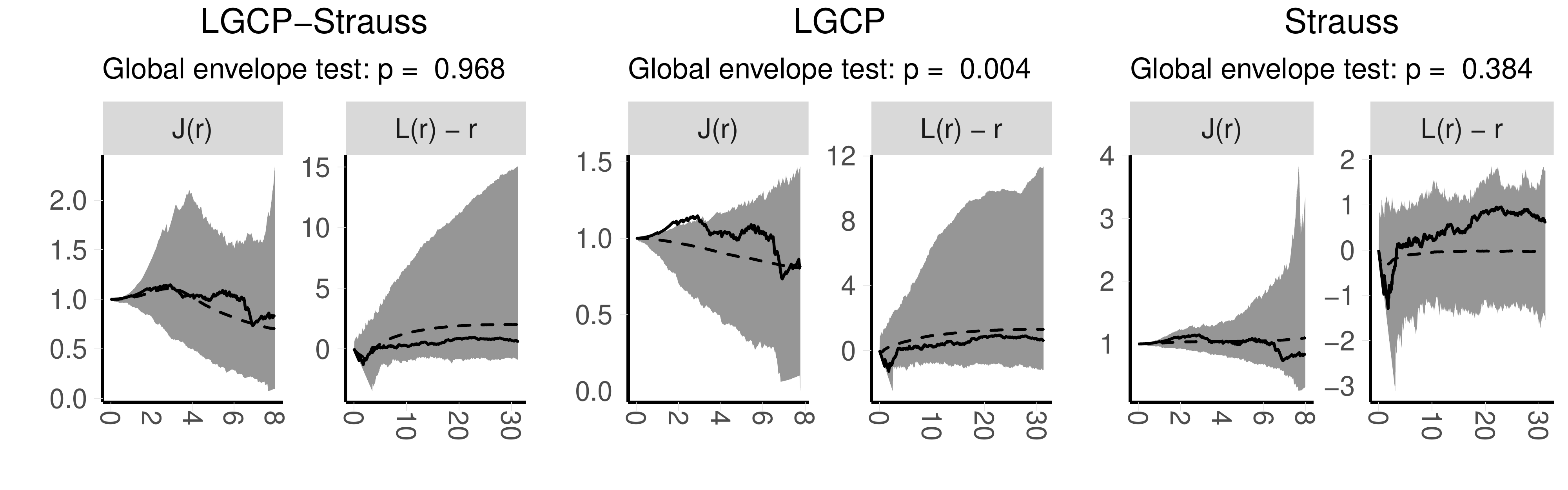}
\caption{Combined global envelopes based on the empirical $J$- and $L$-function for different fitted models to the splited oak point pattern (as indicated at the top of each plot). The solid curves correspond to the splited oak point pattern and the dashed curves are the means obtained from $1\,000$ posterior predictions. The shaded area indicate a 95\% global envelope based on the extreme rank length. At the top of each plot, the $p$-value of the corresponding global envelope test is stated.}
\label{fig:env_comb_oak}
\end{figure}

We now consider the techniques for model checking from Section~\ref{sec:model_checking}.
The first plot in Figure~\ref{fig:env_comb_oak} shows 95\% combined global envelopes for the fitted LGCP-Strauss process as described in Section~\ref{sec:model_checking}. The overall behaviour of the observed point pattern seems to be captured well by the LGCP-Strauss process, and the $p$-value is very high. For comparison, Figure~\ref{fig:env_comb_oak} also shows the corresponding $95\%$ envelopes for an LGCP and a Strauss process model fitted with the ABC procedure in Algorithm~\ref{algo:ABC_procedure}. The combined global envelopes indicate that the LGCP model provides a poor fit to data, but the Strauss process model also fits well. However, the $p$-value is lower than the corresponding $p$-value for the LGCP-Strauss process, indicating that the later may provide a better fit.

When we used the method of ABC-RF, the selected model is an LGCP-Strauss process and the approximate posterior probability is 0.74 showing a relatively high confidence in the chosen model.

All things considered, it seems that the fitted LGCP-Strauss process captures the behaviour of the splited oak point pattern very well, that it provides a much better fit than the LGCP process, and that it provides a somewhat better fit than the Strauss process.

\section{Summary and future work}\label{s:final}

We have proposed a novel spatial point process model which enables capturing of regularity through pairwise interactions and aggregation through a Gaussian process realization.  This doubly stochastic spatial point process generalizes both the customary log Gaussian Cox process and the customary Gibbs process.  Because the likelihood is intractable for this model we have developed model fitting through an ABC method.  We have provided both simulation investigation and a real data application in order to reveal the behaviour of process realizations and also our ability to fit the model and do full inference for given point pattern realizations.

Future work may compare the quality of ABC inference with more traditional MCMC based posterior inference which indeed will be much more time consuming. By this, we mean using a missing data MCMC approach for the case of the LGCP (which is then included into the posterior) or the ancillary variable method by \citet{MoellerEtAl} \citep[see also][]{MurrayEtAl} for the Strauss process. However, this will be  time consuming, especially when we have to make perfect simulations of the Strauss process for the ancillary variable method.
Further future opportunities may consider inhomogeneous point patterns (e.g.\ by including covariate information into the mean function of the Gaussian process), and marked point patterns or so-called multi-type versions of our model \citep[see e.g.][]{textbook}.  Such multi-type modelling may allow attraction or inhibition within types but also introduce attraction or inhibition between types.  A different direction would consider space-time versions.   That is, a realization of the process is seen as a spatial point pattern by \emph{integrating} over a window of time.

\section*{Acknowledgements} The research of the first two authors was supported by The Danish Council for Independent Research | Natural Sciences, grant DFF -- 7014-00074 `Statistics for point processes in space and beyond'. The second author was also supported by the `Centre for Stochastic Geometry and Advanced Bioimaging', funded by grant 8721 from the Villum Foundation.

\bibliography{references}

\begin{thebibliography}{}

\bibitem[Baddeley et~al., 2015]{spatstat}
Baddeley, A., Rubak, E., \& Turner, R. (2015).
\newblock {\em Spatial Point Patterns: Methodology and Applications with {R}}.
\newblock Boca Raton: Chapman and Hall/CRC Press.

\bibitem[Baddeley et~al., 2013]{BaddeleyEtAl}
Baddeley, A., Turner, R., Mateu, J., \& Bevan, A. (2013).
\newblock Hybrids of {G}ibbs point process models and their implementation.
\newblock {\em Journal of Statistical Software}, 55, 1--43.

\bibitem[Beaumont, 2010]{ABCoverview}
Beaumont, M.~A. (2010).
\newblock Approximate {B}ayesian computation in evolution and ecology.
\newblock {\em Annual review of ecology, evolution, and systematics}, 41,
  379--406.

\bibitem[Bernton et~al., 2019]{ABCWasserstein}
Bernton, E., Jacob, P.~E., Gerber, M., \& Robert, C.~P. (2019).
\newblock Approximate {B}ayesian computation with the {W}asserstein distance.
\newblock {\em arXiv preprint arXiv:1905.03747}.

\bibitem[Berthelsen \& Møller, 2008]{kmjm08}
Berthelsen, K.~K. \& Møller, J. (2008).
\newblock Non-parametric {B}ayesian inference for inhomogeneous {M}arkov point
  processes.
\newblock {\em Australian \& New Zealand Journal of Statistics}, 50, 257--272.

\bibitem[Cox, 1955]{cox55}
Cox, D.~R. (1955).
\newblock Some statistical methods connected with series of events.
\newblock {\em Journal of the Royal Statistical Society: Series B
  (Methodological)}, 17, 129--157.

\bibitem[Fearnhead \& Prangle, 2012]{ABCsemi}
Fearnhead, P. \& Prangle, D. (2012).
\newblock Constructing summary statistics for approximate {B}ayesian
  computation: semi-automatic approximate {B}ayesian computation.
\newblock {\em Journal of the Royal Statistical Society: Series B (Statistical
  Methodology)}, 74, 419--474.

\bibitem[Geyer \& M{\o}ller, 1994]{bdMetropolis}
Geyer, C.~J. \& M{\o}ller, J. (1994).
\newblock Simulation procedures and likelihood inference for spatial point
  processes.
\newblock {\em Scandinavian Journal of Statistics}, 21, 359--373.

\bibitem[Goldstein et~al., 2015]{clustering_regularity_Gibbs}
Goldstein, J., Haran, M., Simeonov, I., Fricks, J., \& Chiaromonte, F. (2015).
\newblock An attraction–repulsion point process model for respiratory
  syncytial virus infections.
\newblock {\em Biometrics}, 71, 376--385.

\bibitem[Hastie et~al., 2015]{lasso}
Hastie, T., Tibshirani, R., \& Wainwright, M. (2015).
\newblock {\em Statistical Learning with Sparsity: {T}he Lasso and
  Generalizations}.
\newblock Boca Raton: Chapman and Hall/CRC.

\bibitem[Jiang et~al., 2018]{ABCKullbackLeibler}
Jiang, B., Wu, T.-Y., \& Wong, W.~H. (2018).
\newblock Approximate {B}ayesian computation with {K}ullback-{L}eibler
  divergence as data discrepancy.
\newblock In {\em International Conference on Artificial Intelligence and
  Statistics}  (pp.\ 1711--1721).

\bibitem[Kelly \& Ripley, 1976]{kelly1976}
Kelly, F.~P. \& Ripley, B.~D. (1976).
\newblock A note on {S}trauss's model for clustering.
\newblock {\em Biometrika}, 63, 357--360.

\bibitem[Kendall \& M{\o}ller, 2000]{KendallMoeller}
Kendall, W. \& M{\o}ller, J. (2000).
\newblock Perfect simulation using dominating processes on ordered spaces, with
  application to locally stable point processes.
\newblock {\em Advances in Applied Probability}, 32, 844--865.

\bibitem[Lavancier \& M{\o}ller, 2016]{clustering_regularity_thin}
Lavancier, F. \& M{\o}ller, J. (2016).
\newblock Modelling aggregation on the large scale and regularity on the small
  scale in spatial point pattern datasets.
\newblock {\em Scandinavian Journal of Statistics}, 43, 587--609.

\bibitem[Meinshausen, 2007]{relaxedlasso}
Meinshausen, N. (2007).
\newblock Relaxed {L}asso.
\newblock {\em Computational Statistics \& Data Analysis}, 52, 374--393.

\bibitem[Møller et~al., 1998]{LGCP}
Møller, J., Syversveen, A.~R., \& Waagepetersen, R.~P. (1998).
\newblock Log {G}aussian {C}ox processes.
\newblock {\em Scandinavian Journal of Statistics}, 25, 451--482.

\bibitem[Møller \& Waagepetersen, 2004]{textbook}
Møller, J. \& Waagepetersen, R.~P. (2004).
\newblock {\em Statistical Inference and Simulation for Spatial Point
  Processes}.
\newblock Boca Raton: Chapman and Hall/CRC.

\bibitem[M{\o}ller et~al., 2006]{MoellerEtAl}
M{\o}ller, J., Pettitt, A., Berthelsen, K., \& Reeves, R. (2006).
\newblock An efficient markov chain monte carlo method for distributions with
  intractable normalising constants.
\newblock {\em Biometrika}, 93, 451--458.

\bibitem[M{\o}ller \& Waagepetersen, 2017]{JMRW2017}
M{\o}ller, J. \& Waagepetersen, R.~P. (2017).
\newblock Some recent developments in statistics for spatial point patterns.
\newblock {\em Annual Review of Statistics and Its Application}, 4, 317--342.

\bibitem[Mrkvi{\v{c}}ka et~al., 2018]{GET2018}
Mrkvi{\v{c}}ka, T., Myllym{\"a}ki, M., J{\'i}lek, M., \& Hahn, U. (2018).
\newblock A one-way {ANOVA} test for functional data with graphical
  interpretation.
\newblock Available at arXiv:1612.03608.

\bibitem[Murray et~al., 2006]{MurrayEtAl}
Murray, I., Ghahramani, Z., \& MacKay, D. J.~C. (2006).
\newblock {MCMC} for doubly-intractable distributions.
\newblock In {\em Proceedings of the 22nd Annual {C}onference on Uncertainty in
  Artificial Intelligence}  (pp.\ 359–366).  AUAI Press.

\bibitem[Myllym{\"a}ki \& Mrkvi{\v{c}}ka, 2019]{GETinR}
Myllym{\"a}ki, M. \& Mrkvi{\v{c}}ka, T. (2019).
\newblock {GET}: Global envelopes in {R}.
\newblock arXiv preprint arXiv:1911.06583.

\bibitem[Myllym{\"a}ki et~al., 2017]{GET2017}
Myllym{\"a}ki, M., Mrkvi{\v{c}}ka, T., Grabarnik, P., Seijo, H., \& Hahn, U.
  (2017).
\newblock Global envelope tests for spatial processes.
\newblock {\em Journal of the Royal Statistical Society: Series B (Statistical
  Methodology)}, 79, 381--404.

\bibitem[Park et~al., 2016]{ABCMMD}
Park, M., Jitkrittum, W., \& Sejdinovic, D. (2016).
\newblock K2-abc: Approximate {B}ayesian computation with kernel embeddings.

\bibitem[P{\'e}lissier \& Goreaud, 2015]{ads}
P{\'e}lissier, R. \& Goreaud, F. (2015).
\newblock {ads} package for {R}: A fast unbiased implementation of the
  {$K$}-function family for studying spatial point patterns in irregular-shaped
  sampling windows.
\newblock {\em Journal of Statistical Software}, 63, 1--18.

\bibitem[Pudlo et~al., 2015]{ABCRF}
Pudlo, P., Marin, J.-M., Estoup, A., Cornuet, J.-M., Gautier, M., \& Robert,
  C.~P. (2015).
\newblock {Reliable ABC model choice via random forests}.
\newblock {\em Bioinformatics}, 32, 859--866.

\bibitem[{R Core Team}, 2019]{R}
{R Core Team} (2019).
\newblock {\em R: A Language and Environment for Statistical Computing}.
\newblock R Foundation for Statistical Computing, Vienna, Austria.

\bibitem[Ripley, 1976]{ripley76}
Ripley, B.~D. (1976).
\newblock The second-order analysis of stationary point processes.
\newblock {\em Journal of Applied Probability}, 13, 255--266.

\bibitem[Ripley, 1977]{ripley77}
Ripley, B.~D. (1977).
\newblock Modelling spatial patterns.
\newblock {\em Journal of the Royal Statistical Society: Series B
  (Methodological)}, 39, 172--192.

\bibitem[Schlather, 1999]{schlather99}
Schlather, M. (1999).
\newblock {\em An introduction to positive definite functions and to
  unconditional simulation of random fields}.
\newblock Technical Report st 99-10, Department of Mathematics and Statistics,
  Lancaster University.

\bibitem[Schlather et~al., 2015]{RF2}
Schlather, M., Malinowski, A., Menck, P.~J., Oesting, M., \& Strokorb, K.
  (2015).
\newblock Analysis, simulation and prediction of multivariate random fields
  with package {RandomFields}.
\newblock {\em Journal of Statistical Software}, 63, 1--25.

\bibitem[Schlather et~al., 2019]{RF1}
Schlather, M., Malinowski, A., Oesting, M., Boecker, D., Strokorb, K., Engelke,
  S., Martini, J., Ballani, F., Moreva, O., Auel, J., Menck, P.~J., Gross, S.,
  Ober, U., Ribeiro, P., Ripley, B.~D., Singleton, R., Pfaff, B., \& {R Core
  Team} (2019).
\newblock {RandomFields}: Simulation and analysis of random fields.
\newblock R package version 3.3.6.

\bibitem[Sheather \& Jones, 1991]{bwselection}
Sheather, S.~J. \& Jones, M.~C. (1991).
\newblock A reliable data-based bandwidth selection method for kernel density
  estimation.
\newblock {\em Journal of the Royal Statistical Society. Series B
  (Methodological)}, 53, 683--690.

\bibitem[Shirota \& Gelfand, 2017]{ABCppAlan}
Shirota, S. \& Gelfand, A.~E. (2017).
\newblock Approximate {B}ayesian computation and model assessment for repulsive
  spatial point processes.
\newblock {\em Journal of Computational and Graphical Statistics}, 26,
  646--657.

\bibitem[Soubeyrand et~al., 2013]{ABCppfunctional}
Soubeyrand, S., Carpentier, F., Guiton, F., \& Klein, E.~K. (2013).
\newblock Approximate {B}ayesian computation with functional statistics.
\newblock {\em Statistical Applications in Genetics and Molecular Biology}, 12,
  17--37.

\bibitem[Stoica et~al., 2017]{ABCshadow}
Stoica, R.~S., Philippe, A., Gregori, P., \& Mateu, J. (2017).
\newblock {ABC} {S}hadow algorithm: {A} tool for statistical analysis of
  spatial patterns.
\newblock {\em Statistics and Computing}, 27, 1225--1238.

\bibitem[Strauss, 1975]{strauss1975}
Strauss, D.~J. (1975).
\newblock A model for clustering.
\newblock {\em Biometrika}, 62, 467--475.

\bibitem[{van Lieshout} \& Baddeley, 1996]{Jfunction}
{van Lieshout}, M. N.~M. \& Baddeley, A.~J. (1996).
\newblock A nonparametric measure of spatial interaction in point patterns.
\newblock {\em Statistica Neerlandica}, 50, 344--361.

\bibitem[Wickham, 2016]{ggplot}
Wickham, H. (2016).
\newblock {\em ggplot2: Elegant Graphics for Data Analysis}.
\newblock Springer-Verlag New York.

\bibitem[Zhang, 2004]{covar_unidentifiable}
Zhang, H. (2004).
\newblock Inconsistent estimation and asymptotically equal interpolations in
  model-based geostatistics.
\newblock {\em Journal of the American Statistical Association}, 99, 250--261.

\end{thebibliography}

\appendix
\section{Trace plots for accessing the burn-in for the simulation algorithm}
\begin{figure}[ht!]
\centering
\includegraphics[scale=0.3]{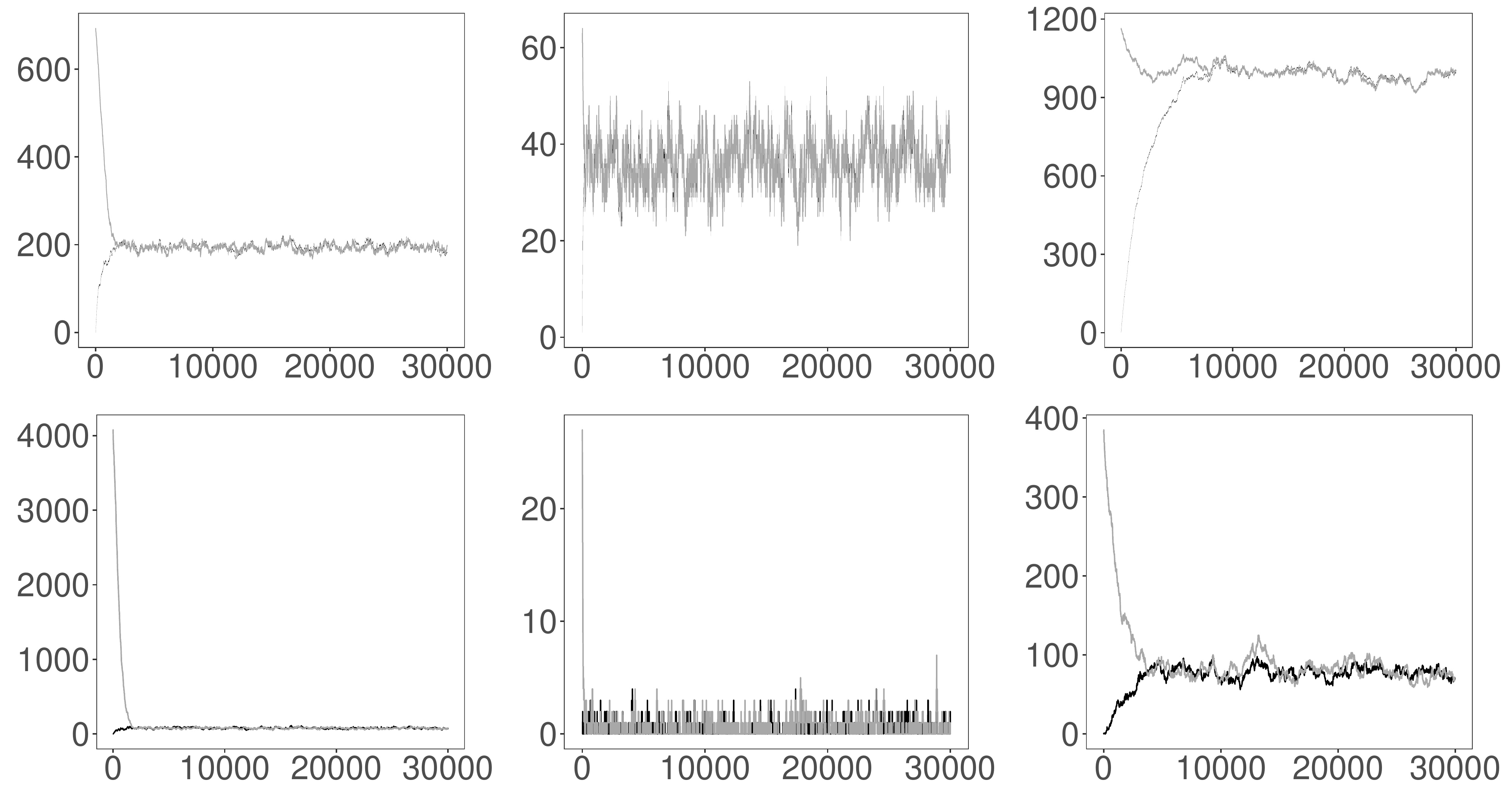}
\caption{Trace plots of the number of points (top) and $R$-close pairs (bottom) for $30\,000$ iterations of the MCMC algorithm for simulating an LGCP-Strauss process on the unit square with parameter vector $\theta$ drawn from the prior distribution $P_1$. Each column of images represent a different sample of $\theta$ and a corresponding realization $\textbf{z}=\{z(u)\}_{u\in W}$ of the GRF $\textbf{Z}$. For each column, the MCMC algorithm was initiated at the empty point pattern (black curves) or a realisation of an inhomogeneous Poisson process with intensity function $\exp\left (z(u)\right )$ (grey curves).}
\label{fig:trace_plots}
\end{figure}

Figure \ref{fig:trace_plots} shows trace plots of the number of points and $R$-close pairs for the MCMC algorithm when simulating an LGCP-Strauss process for different draws of the parameter vector $\theta$ from it's prior distribution $P_1$ which is described in Section~\ref{sec:sim_data_priors}. For each prior sample of $\theta$, a realisation $\textbf{z}$ of the GRF was simulated, and the MCMC algorithm was used to simulate the LGCP-Strauss process given $\textbf{Z}=\textbf{z}$. This analysis was used to choose an appropriate burn-in in Section~\ref{sec:ABCsim}.

\end{document}